\documentclass[]{aastex63}

\usepackage{hyperref}
\usepackage{amsmath}
\usepackage{mathtools}
\usepackage{mathrsfs}
\usepackage{amsbsy}
\usepackage{lineno}
\usepackage{color}\definecolor{AliceBlue}{rgb}{0.94,0.97,1.00}
\definecolor{airforceblue}{rgb}{0.36, 0.54, 0.66}
\definecolor{AntiqueWhite1}{rgb}{1.00,0.94,0.86}
\definecolor{AntiqueWhite2}{rgb}{0.93,0.87,0.80}
\definecolor{AntiqueWhite3}{rgb}{0.80,0.75,0.69}
\definecolor{AntiqueWhite4}{rgb}{0.55,0.51,0.47}
\definecolor{AntiqueWhite}{rgb}{0.98,0.92,0.84}
\definecolor{applered}{rgb}{0.89, 0.02, 0.17}
\definecolor{BlanchedAlmond}{rgb}{1.00,0.92,0.80}
\definecolor{bleudefrance}{rgb}{0.19, 0.55, 0.91}
\definecolor{blue(munsell)}{rgb}{0.0, 0.5, 0.69}
\definecolor{BlueViolet}{rgb}{0.54,0.17,0.89}
\definecolor{CadetBlue1}{rgb}{0.60,0.96,1.00}
\definecolor{CadetBlue2}{rgb}{0.56,0.90,0.93}
\definecolor{CadetBlue3}{rgb}{0.48,0.77,0.80}
\definecolor{CadetBlue4}{rgb}{0.33,0.53,0.55}
\definecolor{CadetBlue}{rgb}{0.37,0.62,0.63}
\definecolor{CornflowerBlue}{rgb}{0.39,0.58,0.93}
\definecolor{DarkBlue}{rgb}{0.00,0.00,0.55}
\definecolor{DarkCyan}{rgb}{0.00,0.55,0.55}
\definecolor{DarkGoldenrod1}{rgb}{1.00,0.73,0.06}
\definecolor{DarkGoldenrod2}{rgb}{0.93,0.68,0.05}
\definecolor{DarkGoldenrod3}{rgb}{0.80,0.58,0.05}
\definecolor{DarkGoldenrod4}{rgb}{0.55,0.40,0.03}
\definecolor{DarkGoldenrod}{rgb}{0.72,0.53,0.04}
\definecolor{DarkGray}{rgb}{0.66,0.66,0.66}
\definecolor{DarkGreen}{rgb}{0.00,0.39,0.00}
\definecolor{DarkGrey}{rgb}{0.66,0.66,0.66}
\definecolor{DarkKhaki}{rgb}{0.74,0.72,0.42}
\definecolor{DarkMagenta}{rgb}{0.55,0.00,0.55}
\definecolor{darkmidnightblue}{rgb}{0.0, 0.2, 0.4}
\definecolor{DarkOliveGreen1}{rgb}{0.79,1.00,0.44}
\definecolor{DarkOliveGreen2}{rgb}{0.74,0.93,0.41}
\definecolor{DarkOliveGreen3}{rgb}{0.64,0.80,0.35}
\definecolor{DarkOliveGreen4}{rgb}{0.43,0.55,0.24}
\definecolor{DarkOliveGreen}{rgb}{0.33,0.42,0.18}
\definecolor{DarkOrange1}{rgb}{1.00,0.50,0.00}
\definecolor{DarkOrange2}{rgb}{0.93,0.46,0.00}
\definecolor{DarkOrange3}{rgb}{0.80,0.40,0.00}
\definecolor{DarkOrange4}{rgb}{0.55,0.27,0.00}
\definecolor{DarkOrange}{rgb}{1.00,0.55,0.00}
\definecolor{DarkOrchid1}{rgb}{0.75,0.24,1.00}
\definecolor{DarkOrchid2}{rgb}{0.70,0.23,0.93}
\definecolor{DarkOrchid3}{rgb}{0.60,0.20,0.80}
\definecolor{DarkOrchid4}{rgb}{0.41,0.13,0.55}
\definecolor{DarkOrchid}{rgb}{0.60,0.20,0.80}
\definecolor{DarkRed}{rgb}{0.55,0.00,0.00}
\definecolor{DarkSalmon}{rgb}{0.91,0.59,0.48}
\definecolor{DarkSeaGreen1}{rgb}{0.76,1.00,0.76}
\definecolor{DarkSeaGreen2}{rgb}{0.71,0.93,0.71}
\definecolor{DarkSeaGreen3}{rgb}{0.61,0.80,0.61}
\definecolor{DarkSeaGreen4}{rgb}{0.41,0.55,0.41}
\definecolor{DarkSeaGreen}{rgb}{0.56,0.74,0.56}
\definecolor{DarkSlateBlue}{rgb}{0.28,0.24,0.55}
\definecolor{DarkSlateGray1}{rgb}{0.59,1.00,1.00}
\definecolor{DarkSlateGray2}{rgb}{0.55,0.93,0.93}
\definecolor{DarkSlateGray3}{rgb}{0.47,0.80,0.80}
\definecolor{DarkSlateGray4}{rgb}{0.32,0.55,0.55}
\definecolor{DarkSlateGray}{rgb}{0.18,0.31,0.31}
\definecolor{DarkSlateGrey}{rgb}{0.18,0.31,0.31}
\definecolor{DarkTurquoise}{rgb}{0.00,0.81,0.82}
\definecolor{DarkViolet}{rgb}{0.58,0.00,0.83}
\definecolor{DeepPink1}{rgb}{1.00,0.08,0.58}
\definecolor{DeepPink2}{rgb}{0.93,0.07,0.54}
\definecolor{DeepPink3}{rgb}{0.80,0.06,0.46}
\definecolor{DeepPink4}{rgb}{0.55,0.04,0.31}
\definecolor{DeepPink}{rgb}{1.00,0.08,0.58}
\definecolor{DeepSkyBlue1}{rgb}{0.00,0.75,1.00}
\definecolor{DeepSkyBlue2}{rgb}{0.00,0.70,0.93}
\definecolor{DeepSkyBlue3}{rgb}{0.00,0.60,0.80}
\definecolor{DeepSkyBlue4}{rgb}{0.00,0.41,0.55}
\definecolor{DeepSkyBlue}{rgb}{0.00,0.75,1.00}
\definecolor{DimGray}{rgb}{0.41,0.41,0.41}
\definecolor{DimGrey}{rgb}{0.41,0.41,0.41}
\definecolor{DodgerBlue1}{rgb}{0.12,0.56,1.00}
\definecolor{DodgerBlue2}{rgb}{0.11,0.53,0.93}
\definecolor{DodgerBlue3}{rgb}{0.09,0.45,0.80}
\definecolor{DodgerBlue4}{rgb}{0.06,0.31,0.55}
\definecolor{DodgerBlue}{rgb}{0.12,0.56,1.00}
\definecolor{FloralWhite}{rgb}{1.00,0.98,0.94}
\definecolor{ForestGreen}{rgb}{0.13,0.55,0.13}
\definecolor{GhostWhite}{rgb}{0.97,0.97,1.00}
\definecolor{GreenYellow}{rgb}{0.68,1.00,0.18}
\definecolor{HotPink1}{rgb}{1.00,0.43,0.71}
\definecolor{HotPink2}{rgb}{0.93,0.42,0.65}
\definecolor{HotPink3}{rgb}{0.80,0.38,0.56}
\definecolor{HotPink4}{rgb}{0.55,0.23,0.38}
\definecolor{HotPink}{rgb}{1.00,0.41,0.71}
\definecolor{IndianRed1}{rgb}{1.00,0.42,0.42}
\definecolor{IndianRed2}{rgb}{0.93,0.39,0.39}
\definecolor{IndianRed3}{rgb}{0.80,0.33,0.33}
\definecolor{IndianRed4}{rgb}{0.55,0.23,0.23}
\definecolor{IndianRed}{rgb}{0.80,0.36,0.36}
\definecolor{LavenderBlush1}{rgb}{1.00,0.94,0.96}
\definecolor{LavenderBlush2}{rgb}{0.93,0.88,0.90}
\definecolor{LavenderBlush3}{rgb}{0.80,0.76,0.77}
\definecolor{LavenderBlush4}{rgb}{0.55,0.51,0.53}
\definecolor{LavenderBlush}{rgb}{1.00,0.94,0.96}
\definecolor{LawnGreen}{rgb}{0.49,0.99,0.00}
\definecolor{LemonChiffon1}{rgb}{1.00,0.98,0.80}
\definecolor{LemonChiffon2}{rgb}{0.93,0.91,0.75}
\definecolor{LemonChiffon3}{rgb}{0.80,0.79,0.65}
\definecolor{LemonChiffon4}{rgb}{0.55,0.54,0.44}
\definecolor{LemonChiffon}{rgb}{1.00,0.98,0.80}
\definecolor{light-blue}{rgb}{0.8,0.85,1}
\definecolor{LightBlue1}{rgb}{0.75,0.94,1.00}
\definecolor{LightBlue2}{rgb}{0.70,0.87,0.93}
\definecolor{LightBlue3}{rgb}{0.60,0.75,0.80}
\definecolor{LightBlue4}{rgb}{0.41,0.51,0.55}
\definecolor{LightBlue}{rgb}{0.68,0.85,0.90}
\definecolor{LightCoral}{rgb}{0.94,0.50,0.50}
\definecolor{LightCyan1}{rgb}{0.88,1.00,1.00}
\definecolor{LightCyan2}{rgb}{0.82,0.93,0.93}
\definecolor{LightCyan3}{rgb}{0.71,0.80,0.80}
\definecolor{LightCyan4}{rgb}{0.48,0.55,0.55}
\definecolor{LightCyan}{rgb}{0.88,1.00,1.00}
\definecolor{LightGoldenrod1}{rgb}{1.00,0.93,0.55}
\definecolor{LightGoldenrod2}{rgb}{0.93,0.86,0.51}
\definecolor{LightGoldenrod3}{rgb}{0.80,0.75,0.44}
\definecolor{LightGoldenrod4}{rgb}{0.55,0.51,0.30}
\definecolor{LightGoldenrodYellow}{rgb}{0.98,0.98,0.82}
\definecolor{LightGoldenrod}{rgb}{0.93,0.87,0.51}
\definecolor{LightGray}{rgb}{0.83,0.83,0.83}
\definecolor{LightGreen}{rgb}{0.56,0.93,0.56}
\definecolor{LightGrey}{rgb}{0.83,0.83,0.83}
\definecolor{LightPink1}{rgb}{1.00,0.68,0.73}
\definecolor{LightPink2}{rgb}{0.93,0.64,0.68}
\definecolor{LightPink3}{rgb}{0.80,0.55,0.58}
\definecolor{LightPink4}{rgb}{0.55,0.37,0.40}
\definecolor{LightPink}{rgb}{1.00,0.71,0.76}
\definecolor{LightSalmon1}{rgb}{1.00,0.63,0.48}
\definecolor{LightSalmon2}{rgb}{0.93,0.58,0.45}
\definecolor{LightSalmon3}{rgb}{0.80,0.51,0.38}
\definecolor{LightSalmon4}{rgb}{0.55,0.34,0.26}
\definecolor{LightSalmon}{rgb}{1.00,0.63,0.48}
\definecolor{LightSeaGreen}{rgb}{0.13,0.70,0.67}
\definecolor{LightSkyBlue1}{rgb}{0.69,0.89,1.00}
\definecolor{LightSkyBlue2}{rgb}{0.64,0.83,0.93}
\definecolor{LightSkyBlue3}{rgb}{0.55,0.71,0.80}
\definecolor{LightSkyBlue4}{rgb}{0.38,0.48,0.55}
\definecolor{LightSkyBlue}{rgb}{0.53,0.81,0.98}
\definecolor{LightSlateBlue}{rgb}{0.52,0.44,1.00}
\definecolor{LightSlateGray}{rgb}{0.47,0.53,0.60}
\definecolor{LightSlateGrey}{rgb}{0.47,0.53,0.60}
\definecolor{LightSteelBlue1}{rgb}{0.79,0.88,1.00}
\definecolor{LightSteelBlue2}{rgb}{0.74,0.82,0.93}
\definecolor{LightSteelBlue3}{rgb}{0.64,0.71,0.80}
\definecolor{LightSteelBlue4}{rgb}{0.43,0.48,0.55}
\definecolor{LightSteelBlue}{rgb}{0.69,0.77,0.87}
\definecolor{LightYellow1}{rgb}{1.00,1.00,0.88}
\definecolor{LightYellow2}{rgb}{0.93,0.93,0.82}
\definecolor{LightYellow3}{rgb}{0.80,0.80,0.71}
\definecolor{LightYellow4}{rgb}{0.55,0.55,0.48}
\definecolor{LightYellow}{rgb}{1.00,1.00,0.88}
\definecolor{LimeGreen}{rgb}{0.20,0.80,0.20}
\definecolor{MediumAquamarine}{rgb}{0.40,0.80,0.67}
\definecolor{MediumBlue}{rgb}{0.00,0.00,0.80}
\definecolor{MediumOrchid1}{rgb}{0.88,0.40,1.00}
\definecolor{MediumOrchid2}{rgb}{0.82,0.37,0.93}
\definecolor{MediumOrchid3}{rgb}{0.71,0.32,0.80}
\definecolor{MediumOrchid4}{rgb}{0.48,0.22,0.55}
\definecolor{MediumOrchid}{rgb}{0.73,0.33,0.83}
\definecolor{MediumPurple1}{rgb}{0.67,0.51,1.00}
\definecolor{MediumPurple2}{rgb}{0.62,0.47,0.93}
\definecolor{MediumPurple3}{rgb}{0.54,0.41,0.80}
\definecolor{MediumPurple4}{rgb}{0.36,0.28,0.55}
\definecolor{MediumPurple}{rgb}{0.58,0.44,0.86}
\definecolor{MediumSeaGreen}{rgb}{0.24,0.70,0.44}
\definecolor{MediumSlateBlue}{rgb}{0.48,0.41,0.93}
\definecolor{MediumSpringGreen}{rgb}{0.00,0.98,0.60}
\definecolor{MediumTurquoise}{rgb}{0.28,0.82,0.80}
\definecolor{MediumVioletRed}{rgb}{0.78,0.08,0.52}
\definecolor{MidnightBlue}{rgb}{0.10,0.10,0.44}
\definecolor{MintCream}{rgb}{0.96,1.00,0.98}
\definecolor{MistyRose1}{rgb}{1.00,0.89,0.88}
\definecolor{MistyRose2}{rgb}{0.93,0.84,0.82}
\definecolor{MistyRose3}{rgb}{0.80,0.72,0.71}
\definecolor{MistyRose4}{rgb}{0.55,0.49,0.48}
\definecolor{MistyRose}{rgb}{1.00,0.89,0.88}
\definecolor{NavajoWhite1}{rgb}{1.00,0.87,0.68}
\definecolor{NavajoWhite2}{rgb}{0.93,0.81,0.63}
\definecolor{NavajoWhite3}{rgb}{0.80,0.70,0.55}
\definecolor{NavajoWhite4}{rgb}{0.55,0.47,0.37}
\definecolor{NavajoWhite}{rgb}{1.00,0.87,0.68}
\definecolor{NavyBlue}{rgb}{0.00,0.00,0.50}
\definecolor{OldLace}{rgb}{0.99,0.96,0.90}
\definecolor{OliveDrab1}{rgb}{0.75,1.00,0.24}
\definecolor{OliveDrab2}{rgb}{0.70,0.93,0.23}
\definecolor{OliveDrab3}{rgb}{0.60,0.80,0.20}
\definecolor{OliveDrab4}{rgb}{0.41,0.55,0.13}
\definecolor{OliveDrab}{rgb}{0.42,0.56,0.14}
\definecolor{OrangeRed1}{rgb}{1.00,0.27,0.00}
\definecolor{OrangeRed2}{rgb}{0.93,0.25,0.00}
\definecolor{OrangeRed3}{rgb}{0.80,0.22,0.00}
\definecolor{OrangeRed4}{rgb}{0.55,0.15,0.00}
\definecolor{OrangeRed}{rgb}{1.00,0.27,0.00}
\definecolor{PaleGoldenrod}{rgb}{0.93,0.91,0.67}
\definecolor{PaleGreen1}{rgb}{0.60,1.00,0.60}
\definecolor{PaleGreen2}{rgb}{0.56,0.93,0.56}
\definecolor{PaleGreen3}{rgb}{0.49,0.80,0.49}
\definecolor{PaleGreen4}{rgb}{0.33,0.55,0.33}
\definecolor{PaleGreen}{rgb}{0.60,0.98,0.60}
\definecolor{PaleTurquoise1}{rgb}{0.73,1.00,1.00}
\definecolor{PaleTurquoise2}{rgb}{0.68,0.93,0.93}
\definecolor{PaleTurquoise3}{rgb}{0.59,0.80,0.80}
\definecolor{PaleTurquoise4}{rgb}{0.40,0.55,0.55}
\definecolor{PaleTurquoise}{rgb}{0.69,0.93,0.93}
\definecolor{PaleVioletRed1}{rgb}{1.00,0.51,0.67}
\definecolor{PaleVioletRed2}{rgb}{0.93,0.47,0.62}
\definecolor{PaleVioletRed3}{rgb}{0.80,0.41,0.54}
\definecolor{PaleVioletRed4}{rgb}{0.55,0.28,0.36}
\definecolor{PaleVioletRed}{rgb}{0.86,0.44,0.58}
\definecolor{PapayaWhip}{rgb}{1.00,0.94,0.84}
\definecolor{PeachPuff1}{rgb}{1.00,0.85,0.73}
\definecolor{PeachPuff2}{rgb}{0.93,0.80,0.68}
\definecolor{PeachPuff3}{rgb}{0.80,0.69,0.58}
\definecolor{PeachPuff4}{rgb}{0.55,0.47,0.40}
\definecolor{PeachPuff}{rgb}{1.00,0.85,0.73}
\definecolor{PowderBlue}{rgb}{0.69,0.88,0.90}
\definecolor{RosyBrown1}{rgb}{1.00,0.76,0.76}
\definecolor{RosyBrown2}{rgb}{0.93,0.71,0.71}
\definecolor{RosyBrown3}{rgb}{0.80,0.61,0.61}
\definecolor{RosyBrown4}{rgb}{0.55,0.41,0.41}
\definecolor{RosyBrown}{rgb}{0.74,0.56,0.56}
\definecolor{RoyalBlue1}{rgb}{0.28,0.46,1.00}
\definecolor{RoyalBlue2}{rgb}{0.26,0.43,0.93}
\definecolor{RoyalBlue3}{rgb}{0.23,0.37,0.80}
\definecolor{RoyalBlue4}{rgb}{0.15,0.25,0.55}
\definecolor{RoyalBlue}{rgb}{0.25,0.41,0.88}
\definecolor{SaddleBrown}{rgb}{0.55,0.27,0.07}
\definecolor{SandyBrown}{rgb}{0.96,0.64,0.38}
\definecolor{SeaGreen1}{rgb}{0.33,1.00,0.62}
\definecolor{SeaGreen2}{rgb}{0.31,0.93,0.58}
\definecolor{SeaGreen3}{rgb}{0.26,0.80,0.50}
\definecolor{SeaGreen4}{rgb}{0.18,0.55,0.34}
\definecolor{SeaGreen}{rgb}{0.18,0.55,0.34}
\definecolor{SkyBlue1}{rgb}{0.53,0.81,1.00}
\definecolor{SkyBlue2}{rgb}{0.49,0.75,0.93}
\definecolor{SkyBlue3}{rgb}{0.42,0.65,0.80}
\definecolor{SkyBlue4}{rgb}{0.29,0.44,0.55}
\definecolor{SkyBlue}{rgb}{0.53,0.81,0.92}
\definecolor{SlateBlue1}{rgb}{0.51,0.44,1.00}
\definecolor{SlateBlue2}{rgb}{0.48,0.40,0.93}
\definecolor{SlateBlue3}{rgb}{0.41,0.35,0.80}
\definecolor{SlateBlue4}{rgb}{0.28,0.24,0.55}
\definecolor{SlateBlue}{rgb}{0.42,0.35,0.80}
\definecolor{SlateGray1}{rgb}{0.78,0.89,1.00}
\definecolor{SlateGray2}{rgb}{0.73,0.83,0.93}
\definecolor{SlateGray3}{rgb}{0.62,0.71,0.80}
\definecolor{SlateGray4}{rgb}{0.42,0.48,0.55}
\definecolor{SlateGray}{rgb}{0.44,0.50,0.56}
\definecolor{SlateGrey}{rgb}{0.44,0.50,0.56}
\definecolor{SpringGreen1}{rgb}{0.00,1.00,0.50}
\definecolor{SpringGreen2}{rgb}{0.00,0.93,0.46}
\definecolor{SpringGreen3}{rgb}{0.00,0.80,0.40}
\definecolor{SpringGreen4}{rgb}{0.00,0.55,0.27}
\definecolor{SpringGreen}{rgb}{0.00,1.00,0.50}
\definecolor{steelblue}{rgb}{0.27, 0.51, 0.71}
\definecolor{SteelBlue1}{rgb}{0.39,0.72,1.00}
\definecolor{SteelBlue2}{rgb}{0.36,0.67,0.93}
\definecolor{SteelBlue3}{rgb}{0.31,0.58,0.80}
\definecolor{SteelBlue4}{rgb}{0.21,0.39,0.55}
\definecolor{SteelBlue}{rgb}{0.27,0.51,0.71}
\definecolor{tealblue}{rgb}{0.21, 0.46, 0.53}
\definecolor{VioletRed1}{rgb}{1.00,0.24,0.59}
\definecolor{VioletRed2}{rgb}{0.93,0.23,0.55}
\definecolor{VioletRed3}{rgb}{0.80,0.20,0.47}
\definecolor{VioletRed4}{rgb}{0.55,0.13,0.32}
\definecolor{VioletRed}{rgb}{0.82,0.13,0.56}
\definecolor{WhiteSmoke}{rgb}{0.96,0.96,0.96}
\definecolor{yaleblue}{rgb}{0.06, 0.3, 0.57}
\definecolor{YellowGreen}{rgb}{0.60,0.80,0.20}
\definecolor{aliceblue}{rgb}{0.94,0.97,1.00}
\definecolor{antiquewhite}{rgb}{0.98,0.92,0.84}
\definecolor{aquamarine1}{rgb}{0.50,1.00,0.83}
\definecolor{aquamarine2}{rgb}{0.46,0.93,0.78}
\definecolor{aquamarine3}{rgb}{0.40,0.80,0.67}
\definecolor{aquamarine4}{rgb}{0.27,0.55,0.45}
\definecolor{aquamarine}{rgb}{0.50,1.00,0.83}
\definecolor{azure}{rgb}{0.0, 0.5, 1.0}
\definecolor{azure1}{rgb}{0.94,1.00,1.00}
\definecolor{azure2}{rgb}{0.88,0.93,0.93}
\definecolor{azure3}{rgb}{0.76,0.80,0.80}
\definecolor{azure4}{rgb}{0.51,0.55,0.55}
\definecolor{beige}{rgb}{0.96,0.96,0.86}
\definecolor{bisque1}{rgb}{1.00,0.89,0.77}
\definecolor{bisque2}{rgb}{0.93,0.84,0.72}
\definecolor{bisque3}{rgb}{0.80,0.72,0.62}
\definecolor{bisque4}{rgb}{0.55,0.49,0.42}
\definecolor{bisque}{rgb}{1.00,0.89,0.77}
\definecolor{black}{rgb}{0.00,0.00,0.00}
\definecolor{blanchedalmond}{rgb}{1.00,0.92,0.80}
\definecolor{blue1}{rgb}{0.00,0.00,1.00}
\definecolor{blue2}{rgb}{0.00,0.00,0.93}
\definecolor{blue3}{rgb}{0.00,0.00,0.80}
\definecolor{blue4}{rgb}{0.00,0.00,0.55}
\definecolor{blueviolet}{rgb}{0.54,0.17,0.89}
\definecolor{blue}{rgb}{0.00,0.00,1.00}
\definecolor{brown1}{rgb}{1.00,0.25,0.25}
\definecolor{brown2}{rgb}{0.93,0.23,0.23}
\definecolor{brown3}{rgb}{0.80,0.20,0.20}
\definecolor{brown4}{rgb}{0.55,0.14,0.14}
\definecolor{brown}{rgb}{0.65,0.16,0.16}
\definecolor{burlywood1}{rgb}{1.00,0.83,0.61}
\definecolor{burlywood2}{rgb}{0.93,0.77,0.57}
\definecolor{burlywood3}{rgb}{0.80,0.67,0.49}
\definecolor{burlywood4}{rgb}{0.55,0.45,0.33}
\definecolor{burlywood}{rgb}{0.87,0.72,0.53}
\definecolor{cadetblue}{rgb}{0.37,0.62,0.63}
\definecolor{chartreuse1}{rgb}{0.50,1.00,0.00}
\definecolor{chartreuse2}{rgb}{0.46,0.93,0.00}
\definecolor{chartreuse3}{rgb}{0.40,0.80,0.00}
\definecolor{chartreuse4}{rgb}{0.27,0.55,0.00}
\definecolor{chartreuse}{rgb}{0.50,1.00,0.00}
\definecolor{chocolate1}{rgb}{1.00,0.50,0.14}
\definecolor{chocolate2}{rgb}{0.93,0.46,0.13}
\definecolor{chocolate3}{rgb}{0.80,0.40,0.11}
\definecolor{chocolate4}{rgb}{0.55,0.27,0.07}
\definecolor{chocolate}{rgb}{0.82,0.41,0.12}
\definecolor{coral1}{rgb}{1.00,0.45,0.34}
\definecolor{coral2}{rgb}{0.93,0.42,0.31}
\definecolor{coral3}{rgb}{0.80,0.36,0.27}
\definecolor{coral4}{rgb}{0.55,0.24,0.18}
\definecolor{coral}{rgb}{1.00,0.50,0.31}
\definecolor{cornflowerblue}{rgb}{0.39,0.58,0.93}
\definecolor{cornsilk1}{rgb}{1.00,0.97,0.86}
\definecolor{cornsilk2}{rgb}{0.93,0.91,0.80}
\definecolor{cornsilk3}{rgb}{0.80,0.78,0.69}
\definecolor{cornsilk4}{rgb}{0.55,0.53,0.47}
\definecolor{cornsilk}{rgb}{1.00,0.97,0.86}
\definecolor{cyan1}{rgb}{0.00,1.00,1.00}
\definecolor{cyan2}{rgb}{0.00,0.93,0.93}
\definecolor{cyan3}{rgb}{0.00,0.80,0.80}
\definecolor{cyan4}{rgb}{0.00,0.55,0.55}
\definecolor{cyan}{rgb}{0.00,1.00,1.00}
\definecolor{darkblue}{rgb}{0.00,0.00,0.55}
\definecolor{darkcyan}{rgb}{0.00,0.55,0.55}
\definecolor{darkgoldenrod}{rgb}{0.72,0.53,0.04}
\definecolor{darkgray}{rgb}{0.66,0.66,0.66}
\definecolor{darkgreen}{rgb}{0.00,0.39,0.00}
\definecolor{darkgrey}{rgb}{0.66,0.66,0.66}
\definecolor{darkkhaki}{rgb}{0.74,0.72,0.42}
\definecolor{darkmagenta}{rgb}{0.55,0.00,0.55}
\definecolor{darkolive}{rgb}{0.33,0.42,0.18}
\definecolor{darkorange}{rgb}{1.00,0.55,0.00}
\definecolor{darkorchid}{rgb}{0.60,0.20,0.80}
\definecolor{darkred}{rgb}{0.55,0.00,0.00}
\definecolor{darksalmon}{rgb}{0.91,0.59,0.48}
\definecolor{darksea}{rgb}{0.56,0.74,0.56}
\definecolor{darkslate}{rgb}{0.18,0.31,0.31}
\definecolor{darkslate}{rgb}{0.18,0.31,0.31}
\definecolor{darkslate}{rgb}{0.28,0.24,0.55}
\definecolor{darkturquoise}{rgb}{0.00,0.81,0.82}
\definecolor{darkviolet}{rgb}{0.58,0.00,0.83}
\definecolor{deeppink}{rgb}{1.00,0.08,0.58}
\definecolor{deepsky}{rgb}{0.00,0.75,1.00}
\definecolor{dimgray}{rgb}{0.41,0.41,0.41}
\definecolor{dimgrey}{rgb}{0.41,0.41,0.41}
\definecolor{dodgerblue}{rgb}{0.12,0.56,1.00}
\definecolor{firebrick1}{rgb}{1.00,0.19,0.19}
\definecolor{firebrick2}{rgb}{0.93,0.17,0.17}
\definecolor{firebrick3}{rgb}{0.80,0.15,0.15}
\definecolor{firebrick4}{rgb}{0.55,0.10,0.10}
\definecolor{firebrick}{rgb}{0.70,0.13,0.13}
\definecolor{floralwhite}{rgb}{1.00,0.98,0.94}
\definecolor{forestgreen}{rgb}{0.13,0.55,0.13}
\definecolor{gainsboro}{rgb}{0.86,0.86,0.86}
\definecolor{ghostwhite}{rgb}{0.97,0.97,1.00}
\definecolor{gold1}{rgb}{1.00,0.84,0.00}
\definecolor{gold2}{rgb}{0.93,0.79,0.00}
\definecolor{gold3}{rgb}{0.80,0.68,0.00}
\definecolor{gold4}{rgb}{0.55,0.46,0.00}
\definecolor{goldenrod1}{rgb}{1.00,0.76,0.15}
\definecolor{goldenrod2}{rgb}{0.93,0.71,0.13}
\definecolor{goldenrod3}{rgb}{0.80,0.61,0.11}
\definecolor{goldenrod4}{rgb}{0.55,0.41,0.08}
\definecolor{goldenrod}{rgb}{0.85,0.65,0.13}
\definecolor{gold}{rgb}{1.00,0.84,0.00}
\definecolor{gray0}{rgb}{0.00,0.00,0.00}
\definecolor{gray100}{rgb}{1.00,1.00,1.00}
\definecolor{gray10}{rgb}{0.10,0.10,0.10}
\definecolor{gray11}{rgb}{0.11,0.11,0.11}
\definecolor{gray12}{rgb}{0.12,0.12,0.12}
\definecolor{gray13}{rgb}{0.13,0.13,0.13}
\definecolor{gray14}{rgb}{0.14,0.14,0.14}
\definecolor{gray15}{rgb}{0.15,0.15,0.15}
\definecolor{gray16}{rgb}{0.16,0.16,0.16}
\definecolor{gray17}{rgb}{0.17,0.17,0.17}
\definecolor{gray18}{rgb}{0.18,0.18,0.18}
\definecolor{gray19}{rgb}{0.19,0.19,0.19}
\definecolor{gray1}{rgb}{0.01,0.01,0.01}
\definecolor{gray20}{rgb}{0.20,0.20,0.20}
\definecolor{gray21}{rgb}{0.21,0.21,0.21}
\definecolor{gray22}{rgb}{0.22,0.22,0.22}
\definecolor{gray23}{rgb}{0.23,0.23,0.23}
\definecolor{gray24}{rgb}{0.24,0.24,0.24}
\definecolor{gray25}{rgb}{0.25,0.25,0.25}
\definecolor{gray26}{rgb}{0.26,0.26,0.26}
\definecolor{gray27}{rgb}{0.27,0.27,0.27}
\definecolor{gray28}{rgb}{0.28,0.28,0.28}
\definecolor{gray29}{rgb}{0.29,0.29,0.29}
\definecolor{gray2}{rgb}{0.02,0.02,0.02}
\definecolor{gray30}{rgb}{0.30,0.30,0.30}
\definecolor{gray31}{rgb}{0.31,0.31,0.31}
\definecolor{gray32}{rgb}{0.32,0.32,0.32}
\definecolor{gray33}{rgb}{0.33,0.33,0.33}
\definecolor{gray34}{rgb}{0.34,0.34,0.34}
\definecolor{gray35}{rgb}{0.35,0.35,0.35}
\definecolor{gray36}{rgb}{0.36,0.36,0.36}
\definecolor{gray37}{rgb}{0.37,0.37,0.37}
\definecolor{gray38}{rgb}{0.38,0.38,0.38}
\definecolor{gray39}{rgb}{0.39,0.39,0.39}
\definecolor{gray3}{rgb}{0.03,0.03,0.03}
\definecolor{gray40}{rgb}{0.40,0.40,0.40}
\definecolor{gray41}{rgb}{0.41,0.41,0.41}
\definecolor{gray42}{rgb}{0.42,0.42,0.42}
\definecolor{gray43}{rgb}{0.43,0.43,0.43}
\definecolor{gray44}{rgb}{0.44,0.44,0.44}
\definecolor{gray45}{rgb}{0.45,0.45,0.45}
\definecolor{gray46}{rgb}{0.46,0.46,0.46}
\definecolor{gray47}{rgb}{0.47,0.47,0.47}
\definecolor{gray48}{rgb}{0.48,0.48,0.48}
\definecolor{gray49}{rgb}{0.49,0.49,0.49}
\definecolor{gray4}{rgb}{0.04,0.04,0.04}
\definecolor{gray50}{rgb}{0.50,0.50,0.50}
\definecolor{gray51}{rgb}{0.51,0.51,0.51}
\definecolor{gray52}{rgb}{0.52,0.52,0.52}
\definecolor{gray53}{rgb}{0.53,0.53,0.53}
\definecolor{gray54}{rgb}{0.54,0.54,0.54}
\definecolor{gray55}{rgb}{0.55,0.55,0.55}
\definecolor{gray56}{rgb}{0.56,0.56,0.56}
\definecolor{gray57}{rgb}{0.57,0.57,0.57}
\definecolor{gray58}{rgb}{0.58,0.58,0.58}
\definecolor{gray59}{rgb}{0.59,0.59,0.59}
\definecolor{gray5}{rgb}{0.05,0.05,0.05}
\definecolor{gray60}{rgb}{0.60,0.60,0.60}
\definecolor{gray61}{rgb}{0.61,0.61,0.61}
\definecolor{gray62}{rgb}{0.62,0.62,0.62}
\definecolor{gray63}{rgb}{0.63,0.63,0.63}
\definecolor{gray64}{rgb}{0.64,0.64,0.64}
\definecolor{gray65}{rgb}{0.65,0.65,0.65}
\definecolor{gray66}{rgb}{0.66,0.66,0.66}
\definecolor{gray67}{rgb}{0.67,0.67,0.67}
\definecolor{gray68}{rgb}{0.68,0.68,0.68}
\definecolor{gray69}{rgb}{0.69,0.69,0.69}
\definecolor{gray6}{rgb}{0.06,0.06,0.06}
\definecolor{gray70}{rgb}{0.70,0.70,0.70}
\definecolor{gray71}{rgb}{0.71,0.71,0.71}
\definecolor{gray72}{rgb}{0.72,0.72,0.72}
\definecolor{gray73}{rgb}{0.73,0.73,0.73}
\definecolor{gray74}{rgb}{0.74,0.74,0.74}
\definecolor{gray75}{rgb}{0.75,0.75,0.75}
\definecolor{gray76}{rgb}{0.76,0.76,0.76}
\definecolor{gray77}{rgb}{0.77,0.77,0.77}
\definecolor{gray78}{rgb}{0.78,0.78,0.78}
\definecolor{gray79}{rgb}{0.79,0.79,0.79}
\definecolor{gray7}{rgb}{0.07,0.07,0.07}
\definecolor{gray80}{rgb}{0.80,0.80,0.80}
\definecolor{gray81}{rgb}{0.81,0.81,0.81}
\definecolor{gray82}{rgb}{0.82,0.82,0.82}
\definecolor{gray83}{rgb}{0.83,0.83,0.83}
\definecolor{gray84}{rgb}{0.84,0.84,0.84}
\definecolor{gray85}{rgb}{0.85,0.85,0.85}
\definecolor{gray86}{rgb}{0.86,0.86,0.86}
\definecolor{gray87}{rgb}{0.87,0.87,0.87}
\definecolor{gray88}{rgb}{0.88,0.88,0.88}
\definecolor{gray89}{rgb}{0.89,0.89,0.89}
\definecolor{gray8}{rgb}{0.08,0.08,0.08}
\definecolor{gray90}{rgb}{0.90,0.90,0.90}
\definecolor{gray91}{rgb}{0.91,0.91,0.91}
\definecolor{gray92}{rgb}{0.92,0.92,0.92}
\definecolor{gray93}{rgb}{0.93,0.93,0.93}
\definecolor{gray94}{rgb}{0.94,0.94,0.94}
\definecolor{gray95}{rgb}{0.95,0.95,0.95}
\definecolor{gray96}{rgb}{0.96,0.96,0.96}
\definecolor{gray97}{rgb}{0.97,0.97,0.97}
\definecolor{gray98}{rgb}{0.98,0.98,0.98}
\definecolor{gray99}{rgb}{0.99,0.99,0.99}
\definecolor{gray9}{rgb}{0.09,0.09,0.09}
\definecolor{gray}{rgb}{0.75,0.75,0.75}
\definecolor{green1}{rgb}{0.00,1.00,0.00}
\definecolor{green2}{rgb}{0.00,0.93,0.00}
\definecolor{green3}{rgb}{0.00,0.80,0.00}
\definecolor{green4}{rgb}{0.00,0.55,0.00}
\definecolor{greenyellow}{rgb}{0.68,1.00,0.18}
\definecolor{green}{rgb}{0.00,1.00,0.00}
\definecolor{grey0}{rgb}{0.00,0.00,0.00}
\definecolor{grey100}{rgb}{1.00,1.00,1.00}
\definecolor{grey10}{rgb}{0.10,0.10,0.10}
\definecolor{grey11}{rgb}{0.11,0.11,0.11}
\definecolor{grey12}{rgb}{0.12,0.12,0.12}
\definecolor{grey13}{rgb}{0.13,0.13,0.13}
\definecolor{grey14}{rgb}{0.14,0.14,0.14}
\definecolor{grey15}{rgb}{0.15,0.15,0.15}
\definecolor{grey16}{rgb}{0.16,0.16,0.16}
\definecolor{grey17}{rgb}{0.17,0.17,0.17}
\definecolor{grey18}{rgb}{0.18,0.18,0.18}
\definecolor{grey19}{rgb}{0.19,0.19,0.19}
\definecolor{grey1}{rgb}{0.01,0.01,0.01}
\definecolor{grey20}{rgb}{0.20,0.20,0.20}
\definecolor{grey21}{rgb}{0.21,0.21,0.21}
\definecolor{grey22}{rgb}{0.22,0.22,0.22}
\definecolor{grey23}{rgb}{0.23,0.23,0.23}
\definecolor{grey24}{rgb}{0.24,0.24,0.24}
\definecolor{grey25}{rgb}{0.25,0.25,0.25}
\definecolor{grey26}{rgb}{0.26,0.26,0.26}
\definecolor{grey27}{rgb}{0.27,0.27,0.27}
\definecolor{grey28}{rgb}{0.28,0.28,0.28}
\definecolor{grey29}{rgb}{0.29,0.29,0.29}
\definecolor{grey2}{rgb}{0.02,0.02,0.02}
\definecolor{grey30}{rgb}{0.30,0.30,0.30}
\definecolor{grey31}{rgb}{0.31,0.31,0.31}
\definecolor{grey32}{rgb}{0.32,0.32,0.32}
\definecolor{grey33}{rgb}{0.33,0.33,0.33}
\definecolor{grey34}{rgb}{0.34,0.34,0.34}
\definecolor{grey35}{rgb}{0.35,0.35,0.35}
\definecolor{grey36}{rgb}{0.36,0.36,0.36}
\definecolor{grey37}{rgb}{0.37,0.37,0.37}
\definecolor{grey38}{rgb}{0.38,0.38,0.38}
\definecolor{grey39}{rgb}{0.39,0.39,0.39}
\definecolor{grey3}{rgb}{0.03,0.03,0.03}
\definecolor{grey40}{rgb}{0.40,0.40,0.40}
\definecolor{grey41}{rgb}{0.41,0.41,0.41}
\definecolor{grey42}{rgb}{0.42,0.42,0.42}
\definecolor{grey43}{rgb}{0.43,0.43,0.43}
\definecolor{grey44}{rgb}{0.44,0.44,0.44}
\definecolor{grey45}{rgb}{0.45,0.45,0.45}
\definecolor{grey46}{rgb}{0.46,0.46,0.46}
\definecolor{grey47}{rgb}{0.47,0.47,0.47}
\definecolor{grey48}{rgb}{0.48,0.48,0.48}
\definecolor{grey49}{rgb}{0.49,0.49,0.49}
\definecolor{grey4}{rgb}{0.04,0.04,0.04}
\definecolor{grey50}{rgb}{0.50,0.50,0.50}
\definecolor{grey51}{rgb}{0.51,0.51,0.51}
\definecolor{grey52}{rgb}{0.52,0.52,0.52}
\definecolor{grey53}{rgb}{0.53,0.53,0.53}
\definecolor{grey54}{rgb}{0.54,0.54,0.54}
\definecolor{grey55}{rgb}{0.55,0.55,0.55}
\definecolor{grey56}{rgb}{0.56,0.56,0.56}
\definecolor{grey57}{rgb}{0.57,0.57,0.57}
\definecolor{grey58}{rgb}{0.58,0.58,0.58}
\definecolor{grey59}{rgb}{0.59,0.59,0.59}
\definecolor{grey5}{rgb}{0.05,0.05,0.05}
\definecolor{grey60}{rgb}{0.60,0.60,0.60}
\definecolor{grey61}{rgb}{0.61,0.61,0.61}
\definecolor{grey62}{rgb}{0.62,0.62,0.62}
\definecolor{grey63}{rgb}{0.63,0.63,0.63}
\definecolor{grey64}{rgb}{0.64,0.64,0.64}
\definecolor{grey65}{rgb}{0.65,0.65,0.65}
\definecolor{grey66}{rgb}{0.66,0.66,0.66}
\definecolor{grey67}{rgb}{0.67,0.67,0.67}
\definecolor{grey68}{rgb}{0.68,0.68,0.68}
\definecolor{grey69}{rgb}{0.69,0.69,0.69}
\definecolor{grey6}{rgb}{0.06,0.06,0.06}
\definecolor{grey70}{rgb}{0.70,0.70,0.70}
\definecolor{grey71}{rgb}{0.71,0.71,0.71}
\definecolor{grey72}{rgb}{0.72,0.72,0.72}
\definecolor{grey73}{rgb}{0.73,0.73,0.73}
\definecolor{grey74}{rgb}{0.74,0.74,0.74}
\definecolor{grey75}{rgb}{0.75,0.75,0.75}
\definecolor{grey76}{rgb}{0.76,0.76,0.76}
\definecolor{grey77}{rgb}{0.77,0.77,0.77}
\definecolor{grey78}{rgb}{0.78,0.78,0.78}
\definecolor{grey79}{rgb}{0.79,0.79,0.79}
\definecolor{grey7}{rgb}{0.07,0.07,0.07}
\definecolor{grey80}{rgb}{0.80,0.80,0.80}
\definecolor{grey81}{rgb}{0.81,0.81,0.81}
\definecolor{grey82}{rgb}{0.82,0.82,0.82}
\definecolor{grey83}{rgb}{0.83,0.83,0.83}
\definecolor{grey84}{rgb}{0.84,0.84,0.84}
\definecolor{grey85}{rgb}{0.85,0.85,0.85}
\definecolor{grey86}{rgb}{0.86,0.86,0.86}
\definecolor{grey87}{rgb}{0.87,0.87,0.87}
\definecolor{grey88}{rgb}{0.88,0.88,0.88}
\definecolor{grey89}{rgb}{0.89,0.89,0.89}
\definecolor{grey8}{rgb}{0.08,0.08,0.08}
\definecolor{grey90}{rgb}{0.90,0.90,0.90}
\definecolor{grey91}{rgb}{0.91,0.91,0.91}
\definecolor{grey92}{rgb}{0.92,0.92,0.92}
\definecolor{grey93}{rgb}{0.93,0.93,0.93}
\definecolor{grey94}{rgb}{0.94,0.94,0.94}
\definecolor{grey95}{rgb}{0.95,0.95,0.95}
\definecolor{grey96}{rgb}{0.96,0.96,0.96}
\definecolor{grey97}{rgb}{0.97,0.97,0.97}
\definecolor{grey98}{rgb}{0.98,0.98,0.98}
\definecolor{grey99}{rgb}{0.99,0.99,0.99}
\definecolor{grey9}{rgb}{0.09,0.09,0.09}
\definecolor{grey}{rgb}{0.75,0.75,0.75}
\definecolor{honeydew1}{rgb}{0.94,1.00,0.94}
\definecolor{honeydew2}{rgb}{0.88,0.93,0.88}
\definecolor{honeydew3}{rgb}{0.76,0.80,0.76}
\definecolor{honeydew4}{rgb}{0.51,0.55,0.51}
\definecolor{honeydew}{rgb}{0.94,1.00,0.94}
\definecolor{hotpink}{rgb}{1.00,0.41,0.71}
\definecolor{indianred}{rgb}{0.80,0.36,0.36}
\definecolor{ivory1}{rgb}{1.00,1.00,0.94}
\definecolor{ivory2}{rgb}{0.93,0.93,0.88}
\definecolor{ivory3}{rgb}{0.80,0.80,0.76}
\definecolor{ivory4}{rgb}{0.55,0.55,0.51}
\definecolor{ivory}{rgb}{1.00,1.00,0.94}
\definecolor{khaki1}{rgb}{1.00,0.96,0.56}
\definecolor{khaki2}{rgb}{0.93,0.90,0.52}
\definecolor{khaki3}{rgb}{0.80,0.78,0.45}
\definecolor{khaki4}{rgb}{0.55,0.53,0.31}
\definecolor{khaki}{rgb}{0.94,0.90,0.55}
\definecolor{lavenderblush}{rgb}{1.00,0.94,0.96}
\definecolor{lavender}{rgb}{0.90,0.90,0.98}
\definecolor{lawngreen}{rgb}{0.49,0.99,0.00}
\definecolor{lemonchiffon}{rgb}{1.00,0.98,0.80}
\definecolor{lightblue}{rgb}{0.68,0.85,0.90}
\definecolor{lightcoral}{rgb}{0.94,0.50,0.50}
\definecolor{lightcyan}{rgb}{0.88,1.00,1.00}
\definecolor{lightgoldenrod}{rgb}{0.93,0.87,0.51}
\definecolor{lightgoldenrod}{rgb}{0.98,0.98,0.82}
\definecolor{lightgray}{rgb}{0.83,0.83,0.83}
\definecolor{lightgreen}{rgb}{0.56,0.93,0.56}
\definecolor{lightgrey}{rgb}{0.83,0.83,0.83}
\definecolor{lightpink}{rgb}{1.00,0.71,0.76}
\definecolor{lightsalmon}{rgb}{1.00,0.63,0.48}
\definecolor{lightsea}{rgb}{0.13,0.70,0.67}
\definecolor{lightsky}{rgb}{0.53,0.81,0.98}
\definecolor{lightslate}{rgb}{0.47,0.53,0.60}
\definecolor{lightslate}{rgb}{0.47,0.53,0.60}
\definecolor{lightslate}{rgb}{0.52,0.44,1.00}
\definecolor{lightsteel}{rgb}{0.69,0.77,0.87}
\definecolor{lightyellow}{rgb}{1.00,1.00,0.88}
\definecolor{limegreen}{rgb}{0.20,0.80,0.20}
\definecolor{linen}{rgb}{0.98,0.94,0.90}
\definecolor{magenta1}{rgb}{1.00,0.00,1.00}
\definecolor{magenta2}{rgb}{0.93,0.00,0.93}
\definecolor{magenta3}{rgb}{0.80,0.00,0.80}
\definecolor{magenta4}{rgb}{0.55,0.00,0.55}
\definecolor{magenta}{rgb}{1.00,0.00,1.00}
\definecolor{maroon1}{rgb}{1.00,0.20,0.70}
\definecolor{maroon2}{rgb}{0.93,0.19,0.65}
\definecolor{maroon3}{rgb}{0.80,0.16,0.56}
\definecolor{maroon4}{rgb}{0.55,0.11,0.38}
\definecolor{maroon}{rgb}{0.69,0.19,0.38}
\definecolor{mediumaquamarine}{rgb}{0.40,0.80,0.67}
\definecolor{mediumblue}{rgb}{0.00,0.00,0.80}
\definecolor{mediumorchid}{rgb}{0.73,0.33,0.83}
\definecolor{mediumpurple}{rgb}{0.58,0.44,0.86}
\definecolor{mediumsea}{rgb}{0.24,0.70,0.44}
\definecolor{mediumslate}{rgb}{0.48,0.41,0.93}
\definecolor{mediumspring}{rgb}{0.00,0.98,0.60}
\definecolor{mediumturquoise}{rgb}{0.28,0.82,0.80}
\definecolor{mediumviolet}{rgb}{0.78,0.08,0.52}
\definecolor{midnightblue}{rgb}{0.10,0.10,0.44}
\definecolor{mintcream}{rgb}{0.96,1.00,0.98}
\definecolor{mistyrose}{rgb}{1.00,0.89,0.88}
\definecolor{moccasin}{rgb}{1.00,0.89,0.71}
\definecolor{navajowhite}{rgb}{1.00,0.87,0.68}
\definecolor{navyblue}{rgb}{0.00,0.00,0.50}
\definecolor{navy}{rgb}{0.00,0.00,0.50}
\definecolor{oldlace}{rgb}{0.99,0.96,0.90}
\definecolor{olivedrab}{rgb}{0.42,0.56,0.14}
\definecolor{orange1}{rgb}{1.00,0.65,0.00}
\definecolor{orange2}{rgb}{0.93,0.60,0.00}
\definecolor{orange3}{rgb}{0.80,0.52,0.00}
\definecolor{orange4}{rgb}{0.55,0.35,0.00}
\definecolor{orangered}{rgb}{1.00,0.27,0.00}
\definecolor{orange}{rgb}{1.00,0.65,0.00}
\definecolor{orchid1}{rgb}{1.00,0.51,0.98}
\definecolor{orchid2}{rgb}{0.93,0.48,0.91}
\definecolor{orchid3}{rgb}{0.80,0.41,0.79}
\definecolor{orchid4}{rgb}{0.55,0.28,0.54}
\definecolor{orchid}{rgb}{0.85,0.44,0.84}
\definecolor{palegoldenrod}{rgb}{0.93,0.91,0.67}
\definecolor{palegreen}{rgb}{0.60,0.98,0.60}
\definecolor{paleturquoise}{rgb}{0.69,0.93,0.93}
\definecolor{paleviolet}{rgb}{0.86,0.44,0.58}
\definecolor{papayawhip}{rgb}{1.00,0.94,0.84}
\definecolor{peachpuff}{rgb}{1.00,0.85,0.73}
\definecolor{peru}{rgb}{0.80,0.52,0.25}
\definecolor{pink1}{rgb}{1.00,0.71,0.77}
\definecolor{pink2}{rgb}{0.93,0.66,0.72}
\definecolor{pink3}{rgb}{0.80,0.57,0.62}
\definecolor{pink4}{rgb}{0.55,0.39,0.42}
\definecolor{pink}{rgb}{1.00,0.75,0.80}
\definecolor{plum1}{rgb}{1.00,0.73,1.00}
\definecolor{plum2}{rgb}{0.93,0.68,0.93}
\definecolor{plum3}{rgb}{0.80,0.59,0.80}
\definecolor{plum4}{rgb}{0.55,0.40,0.55}
\definecolor{plum}{rgb}{0.87,0.63,0.87}
\definecolor{powderblue}{rgb}{0.69,0.88,0.90}
\definecolor{purple1}{rgb}{0.61,0.19,1.00}
\definecolor{purple2}{rgb}{0.57,0.17,0.93}
\definecolor{purple3}{rgb}{0.49,0.15,0.80}
\definecolor{purple4}{rgb}{0.33,0.10,0.55}
\definecolor{purple}{rgb}{0.63,0.13,0.94}
\definecolor{red1}{rgb}{1.00,0.00,0.00}
\definecolor{red2}{rgb}{0.93,0.00,0.00}
\definecolor{red3}{rgb}{0.80,0.00,0.00}
\definecolor{red4}{rgb}{0.55,0.00,0.00}
\definecolor{red}{rgb}{1.00,0.00,0.00}
\definecolor{rosybrown}{rgb}{0.74,0.56,0.56}
\definecolor{royalblue}{rgb}{0.25,0.41,0.88}
\definecolor{saddlebrown}{rgb}{0.55,0.27,0.07}
\definecolor{salmon1}{rgb}{1.00,0.55,0.41}
\definecolor{salmon2}{rgb}{0.93,0.51,0.38}
\definecolor{salmon3}{rgb}{0.80,0.44,0.33}
\definecolor{salmon4}{rgb}{0.55,0.30,0.22}
\definecolor{salmon}{rgb}{0.98,0.50,0.45}
\definecolor{sandybrown}{rgb}{0.96,0.64,0.38}
\definecolor{seagreen}{rgb}{0.18,0.55,0.34}
\definecolor{seashell1}{rgb}{1.00,0.96,0.93}
\definecolor{seashell2}{rgb}{0.93,0.90,0.87}
\definecolor{seashell3}{rgb}{0.80,0.77,0.75}
\definecolor{seashell4}{rgb}{0.55,0.53,0.51}
\definecolor{seashell}{rgb}{1.00,0.96,0.93}
\definecolor{sienna1}{rgb}{1.00,0.51,0.28}
\definecolor{sienna2}{rgb}{0.93,0.47,0.26}
\definecolor{sienna3}{rgb}{0.80,0.41,0.22}
\definecolor{sienna4}{rgb}{0.55,0.28,0.15}
\definecolor{sienna}{rgb}{0.63,0.32,0.18}
\definecolor{skyblue}{rgb}{0.53,0.81,0.92}
\definecolor{slateblue}{rgb}{0.42,0.35,0.80}
\definecolor{slategray}{rgb}{0.44,0.50,0.56}
\definecolor{slategrey}{rgb}{0.44,0.50,0.56}
\definecolor{snow1}{rgb}{1.00,0.98,0.98}
\definecolor{snow2}{rgb}{0.93,0.91,0.91}
\definecolor{snow3}{rgb}{0.80,0.79,0.79}
\definecolor{snow4}{rgb}{0.55,0.54,0.54}
\definecolor{snow}{rgb}{1.00,0.98,0.98}
\definecolor{springgreen}{rgb}{0.00,1.00,0.50}
\definecolor{steelblue}{rgb}{0.27,0.51,0.71}
\definecolor{tan1}{rgb}{1.00,0.65,0.31}
\definecolor{tan2}{rgb}{0.93,0.60,0.29}
\definecolor{tan3}{rgb}{0.80,0.52,0.25}
\definecolor{tan4}{rgb}{0.55,0.35,0.17}
\definecolor{tan}{rgb}{0.82,0.71,0.55}
\definecolor{thistle1}{rgb}{1.00,0.88,1.00}
\definecolor{thistle2}{rgb}{0.93,0.82,0.93}
\definecolor{thistle3}{rgb}{0.80,0.71,0.80}
\definecolor{thistle4}{rgb}{0.55,0.48,0.55}
\definecolor{thistle}{rgb}{0.85,0.75,0.85}
\definecolor{tomato1}{rgb}{1.00,0.39,0.28}
\definecolor{tomato2}{rgb}{0.93,0.36,0.26}
\definecolor{tomato3}{rgb}{0.80,0.31,0.22}
\definecolor{tomato4}{rgb}{0.55,0.21,0.15}
\definecolor{tomato}{rgb}{1.00,0.39,0.28}
\definecolor{turquoise1}{rgb}{0.00,0.96,1.00}
\definecolor{turquoise2}{rgb}{0.00,0.90,0.93}
\definecolor{turquoise3}{rgb}{0.00,0.77,0.80}
\definecolor{turquoise4}{rgb}{0.00,0.53,0.55}
\definecolor{turquoise}{rgb}{0.25,0.88,0.82}
\definecolor{violetred}{rgb}{0.82,0.13,0.56}
\definecolor{violet}{rgb}{0.93,0.51,0.93}
\definecolor{wheat1}{rgb}{1.00,0.91,0.73}
\definecolor{wheat2}{rgb}{0.93,0.85,0.68}
\definecolor{wheat3}{rgb}{0.80,0.73,0.59}
\definecolor{wheat4}{rgb}{0.55,0.49,0.40}
\definecolor{wheat}{rgb}{0.96,0.87,0.70}
\definecolor{whitesmoke}{rgb}{0.96,0.96,0.96}
\definecolor{white}{rgb}{1.00,1.00,1.00}
\definecolor{yellow1}{rgb}{1.00,1.00,0.00}
\definecolor{yellow2}{rgb}{0.93,0.93,0.00}
\definecolor{yellow3}{rgb}{0.80,0.80,0.00}
\definecolor{yellow4}{rgb}{0.55,0.55,0.00}
\definecolor{yellowgreen}{rgb}{0.60,0.80,0.20}
\definecolor{yellow}{rgb}{1.00,1.00,0.00}

\newcommand{\jcafont}{
  \color{red}
}
\newcommand{\gskfont}{
  \color{orange}
}
\newcommand{\agefont}{
  \color{purple}
}
\DeclareTextFontCommand{\gsk}{\gskfont}
\DeclareTextFontCommand{\jca}{\jcafont}
\DeclareTextFontCommand{\age}{\agefont}

\newcommand{\sn}[2]{{#1} \times 10^{#2}}

\newcommand{\ionwl}[3]{\ion{#1}{#2}~{#3}~{\AA}}

\newcommand{\radyn}{\texttt{RADYN}}
\renewcommand{\fp}{\texttt{FP}}

\shorttitle{Solar Flares with Suppressed Conduction}
\shortauthors{Allred et al.}

\begin{document}

\title{Solar Flare Heating with Turbulent Suppression of Thermal Conduction}

\correspondingauthor{Joel C. Allred}
\email{joel.c.allred@nasa.gov}

\author{Joel C. Allred}
\affil{NASA Goddard Space Flight Center, Solar Physics Laboratory, Code 671, Greenbelt, MD 20771, USA}

\author{Graham S. Kerr}
\affil{NASA Goddard Space Flight Center, Solar Physics Laboratory, Code 671, Greenbelt, MD 20771, USA}
 	\affil{Department of Physics, Catholic University of America, 620 Michigan Avenue, Northeast, Washington, DC 20064, USA}

\author{A. Gordon Emslie}
\affiliation{Department of Physics \& Astronomy, Western Kentucky University, Bowling Green, KY 42101, USA}

\begin{abstract}
During solar flares plasma is typically heated to very high temperatures, and the resulting redistribution of energy via thermal conduction is a primary mechanism transporting energy throughout the flaring solar atmosphere. The thermal flux is usually modeled using Spitzer's theory, which is based on local Coulomb collisions between the electrons carrying the thermal flux and those in the background. However, often during flares, temperature gradients become sufficiently steep that the collisional mean free path exceeds the temperature gradient scale size, so that thermal conduction becomes inherently non-local. Further, turbulent angular scattering, which is detectable in nonthermal widths of atomic emission lines, can also act to increase the collision frequency and so suppress the heat flux. Recent work by Emslie \& Bian (2018) extended Spitzer's theory of thermal conduction to account for both non-locality and turbulent suppression. We have implemented their theoretical expression for the heat flux (which is a convolution of the Spitzer flux with a kernel function) into the RADYN flare-modeling code and performed a parameter study to understand how the resulting changes in thermal conduction affect flare dynamics and hence the radiation produced. We find that models with reduced heat fluxes predict slower bulk flows, less intense line emission, and longer cooling times. By comparing features of atomic emission lines predicted by the models with Doppler velocities and nonthermal line widths deduced from a particular flare observation, we find that models with suppression factors between 0.3 to 0.5 relative to the Spitzer value best reproduce observed Doppler velocities across emission lines forming over a wide range of temperatures. Interestingly, the model that best matches observed nonthermal line widths has a kappa-type velocity distribution function. 
\end{abstract}

\section{Introduction} \label{sec:intro}
The flare component of a solar eruptive event rapidly heats plasma in the Sun's atmosphere to temperatures often greater than 20~MK \citep[e.g.,][]{2011SSRv..159...19F, 2014ApJ...781...43C, 2019ApJ...881..161M}. The hot flaring corona is thermally connected to the cooler chromosphere ($\lesssim 20$~kK) through a relatively narrow transition region with a steep temperature gradient, and the thermal conductive flux associated with such large temperature gradients is a primary mechanism for transporting energy throughout flaring loops \citep[e.g.,][and references therein]{2001SoPh..204...91A, 2013ApJ...778...68R, 2016A&A...588A.116W, 2020A&A...644A.172W}.  

When the electron-ion collisional mean free path $\lambda_{ei}$ is long compared to the temperature gradient scale size $L$ (i.e., the Knudsen number $\lambda_{ei}/L \gg 1$), then thermal conduction becomes an inherently non-local process. In this regime, \citet{1984PhRvA..30..365C} has demonstrated that thermal conduction is reduced relative to the \cite{1962pfig.book.....S} formalism. Additionally, \citet{2016ApJ...824...78B} have shown that turbulent angular scattering of thermal electrons further reduces the thermal conductivity and, since thermal conduction is a main driver of flare dynamics, the overall evolution of temperatures, densities, and bulk velocities within the loop are all affected by the presence of turbulence.

The presence of turbulence is inferred from the broadening of atomic spectral line profiles and has been detected during flares in numerous previous studies \citep[e.g.,][and references therein]{2003ApJ...582..506L,2006SoPh..234...95D,2011ApJ...740...70M, 2015ApJ...799..218Y,2016A&A...590A..99J, 2018ApJ...864...63P,2020ApJ...905..165R}.  Atomic emission line profiles are therefore sensitive to turbulence both directly (through nonthermal broadening) and indirectly, through the hydrodynamic evolution of flare loops driven in part by a turbulence-modified conduction term. This makes such line profiles ideal diagnostics for studying the role of turbulence in flares. 

\citet[][hereafter EB18]{2018ApJ...865...67E} have developed a new formalism describing the suppression of thermal conduction that includes both non-local and turbulent effects in a single kernel that must be convolved with the local Spitzer heat flux to obtain the actual heat flux.
In this work, we apply this approach to study the suppression of thermal conduction by both non-locality and turbulence, by modeling the radiative hydrodynamic evolution of flare loops to several parameterizations of the turbulence spectrum. In Section~\ref{sec:method}, we present our methodology for performing this parameter study. These models can then be used to predict emission profiles in many atomic lines, and in Section~\ref{sec:discuss} we compare predicted Doppler and nonthermal velocities with those observed during a particular flare. Finally in Sectione~\ref{sec:conc}, we summarize these results and present our conclusions.

\section{Methodology}\label{sec:method}
We model the hydrodynamic response to the heating produced from an injection of nonthermal electrons into a loop using the \radyn\ code \citep{1992ApJ...397L..59C,1995ApJ...440L..29C, 1997ApJ...481..500C, 2005ApJ...630..573A, 2015ApJ...809..104A} combined with the \fp\ particle transport code \citep{2020ApJ...902...16A}. \radyn\ has been used extensively in previous studies of flare heating \citep[e.g.,][]{1999ApJ...521..906A, 2005ApJ...630..573A, 2015SoPh..290.3487K, 2017ApJ...836...12K, 2018ApJ...852...61K, 2022ApJ...928..190K, 2015A&A...578A..72K, 2016ApJ...827...38R, 2016ApJ...827..101K, 2019ApJ...871...23K, 2020ApJ...900...18K,2021ApJ...912..153K, 2017A&A...605A.125S,2018ApJ...862...59B, 2020ApJ...895....6G} and its computational method is described in those works.  Briefly, it solves the 1D equations of hydrodynamics coupled with non-equilibrium atomic level population equations for six levels of H, nine levels of \ion{He}{1} and \ion{He}{2}, and six levels of \ion{Ca}{2} on an adaptive grid. The radiative transfer equation is solved using a 1.5D plane parallel approximation. A key feature of \radyn\ is its ability to model optically-thick, non-LTE, lines and continua for many transitions that are important for energy transport in the chromosphere, where much of the flare energy is deposited.

For this study, we have modified the expression that \radyn's uses to compute thermal conduction to incorporate the formalism for non-local and turbulent conduction presented in EB18. First the local \cite{1962pfig.book.....S} heat flux, $F_S$, is calculated as usual:

\begin{linenomath*} \begin{equation}
F_S = -\kappa_o \, T^{5/2} \, \frac{dT}{dz} \,\,\, ,
\label{eqn:spitzer}
\end{equation} \end{linenomath*}
where $\kappa_o = \sn{1.1}{-6}$ erg K$^{-7/2}$  cm$^{-1}$ s$^{-1}$, $T$ is the plasma temperature (K), and $z$ is the distance (cm) along the loop, measured upward from the photosphere. $F_S$ is then convolved with a kernel, $Q$, that accounts for both non-locality and turbulence, as derived in EB18. 

To model the effect of turbulence on thermal electron transport, EB18 assumed the velocity dependence of the turbulent mean free path, $\lambda_T(v)$, has the form 

\begin{linenomath*} \begin{equation}
\lambda_T(v) = \lambda_{T0} \left( \frac{v}{v_{te}}\right)^\alpha = \frac{\lambda_{ei}}{R} \left( \frac{v}{v_{te}}\right)^\alpha\,\,\, ,
\label{eqn:vturb} 
\end{equation} \end{linenomath*}
with $R = \lambda_{ei}/\lambda_{T0}$. Here

\begin{linenomath*} \begin{equation}
v_{te} = \sqrt{\frac{2 k T}{m_e}}
\end{equation} \end{linenomath*}
is the electron thermal speed and $\lambda_{ei}$ is the electron collisional mean free path at the thermal speed:

\begin{linenomath*} \begin{equation} 
\lambda_{ei} = \frac{m_e^2 \, v_{te}^4}{ 4 \pi e^4 \ln \Lambda \, n} \,\,\, .
\end{equation} \end{linenomath*} 
where $k$ ($\sn{1.38}{-16}$~erg K$^{-1}$) is the Boltzmann constant, $e$ ($\sn{4.8}{-10}$~esu) is the electronic charge, $n$ (cm$^{-3}$) is the background electron density, $m_e$ ($\sn{9.1}{-28}$~g) is the electron mass, and $\ln \Lambda (\simeq 20$) is the Coulomb logarithm.

As shown by EB18, the form of the heat flux in the presence of nonlocal effects and/or turbulence is given by the convolution

\begin{figure}
\plotone{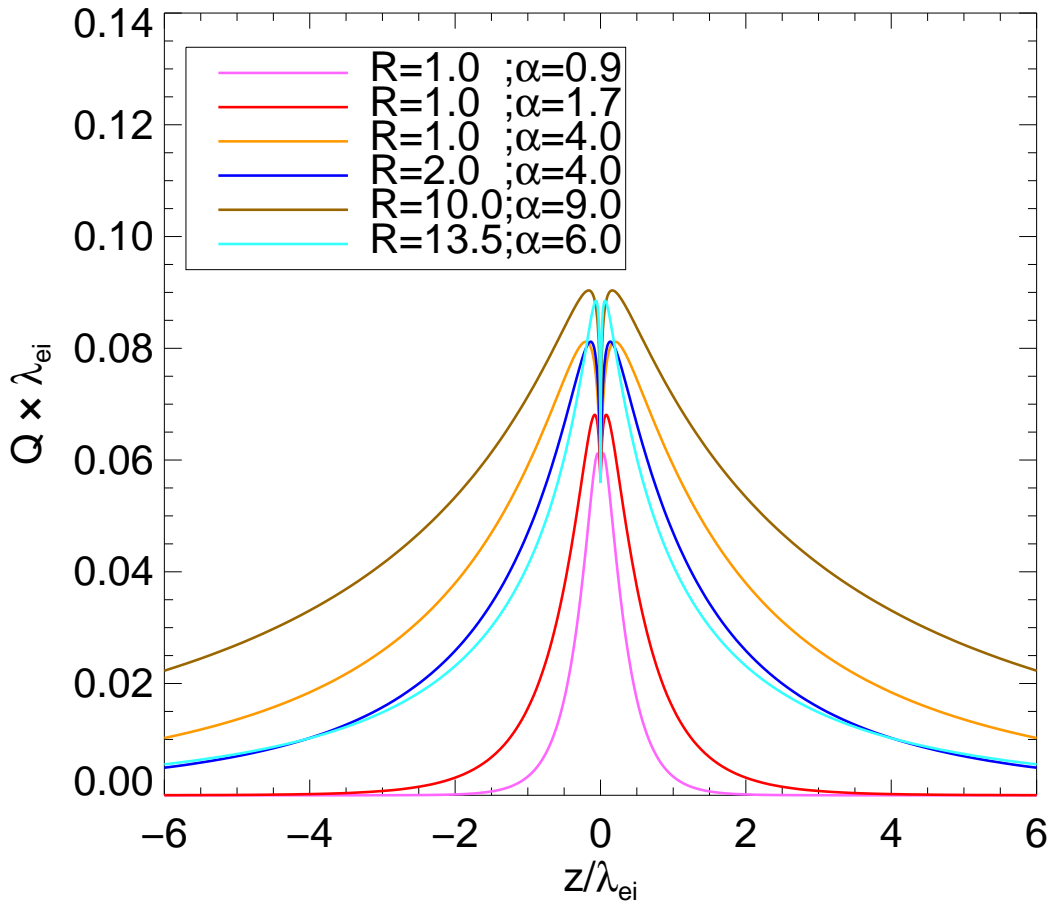}
\caption{The convolution kernel, $Q$ (Equation~(\ref{eqn:q-kernel})), for several values of $R$ and $\alpha$.}
\label{fig:kernel}
\end{figure}

\begin{linenomath*} \begin{equation}
F_c(z) = \int Q\left(z-z'; R, \alpha\right) \, F_S(z') \, dz'\,\,\, , 
\end{equation} \end{linenomath*}
where the integral is done over the full loop length and the kernel function $Q$ is given by (EB18, their Equation~(27))

\begin{equation}\label{eqn:q-kernel}
Q\left(z; R, \alpha\right) = \frac{\sqrt{45}}{60 \, \lambda_{ei}} \int_{0}^{\infty} x^5 \left(x^2 - \frac{5}{2}\right) \exp\left[-\left(x^2 +\frac{\sqrt{45}\left(1+R x^{4-\alpha}\right)}{x^4} \left|\frac{z}{\lambda_{ei}}\right|\right)\right] \, dx \,\,\, .
\end{equation}
$Q$ is thus parameterized by $R$ and $\alpha$, and is plotted in Figure~\ref{fig:kernel} for several values of $R$ and $\alpha$ that have been used in this study.  We note that \radyn\ models half-loops, and by reflecting the \radyn\ half-loop about the apex, we make the assumption that the full loop is symmetric.

The initial (preflare) state of the loop model has an apex temperature and electron density of $3.3$~MK and $\sn{7.5}{9}$~cm$^{-3}$, respectively. The temperature and density as a function of distance are plotted as the $t=0$~s (black line) in Figure~\ref{fig:loopevol}. This state is obtained by evolving the radiative hydrodynamic equations with a coronal heating term as a source of energy at the top of the loop until a state of hydrostatic equilibrium is obtained, as described in \citet{2015ApJ...809..104A}. In doing this relaxation to equilibrium, the unsuppressed Spitzer heat flux is used, so all models presented in \S\ref{sec:discuss} (listed in Table~\ref{tab:params}) have the same initial state.

We then inject a beam of nonthermal electrons at the apex of a model flux loop and track their transport until their eventual thermalization in the footpoints, typically in the chromosphere. For each case studied here, we inject the same nonthermal electron distribution, which is characterized by a power-law with cutoff energy, $E_c = 12$~keV, spectral index, $\delta = 7.6$, and energy flux $F = \sn{5}{10}$~erg~cm$^{-2}$~s$^{-1}$. This electron beam distribution was chosen to correspond to that which was inferred at the peak of a C-class flare studied by \citet[][hereafter MD09]{2009ApJ...699..968M} and \citet[][hereafter M11]{2011ApJ...740...70M} and will aid in comparing their observations with our model predictions presented in the subsequent section. In each case the beam is injected for $\Delta t = 20$~s as estimated from observed impulsive rise times. 

The transport of injected electrons is modeled using \fp, which we have configured to run in a 1D mode, meaning all electrons are injected along the magnetic field line axis and pitch-angle diffusion due to Coulomb collisions is neglected. As a test, we have run cases in which pitch-angle diffusion is included and found it makes little difference to the beam heating rate, which is the important quantity for these simulations. We recognize that diffusive scattering of the nonthermal electrons by turbulence could alter their energy deposition rate, especially if turbulent scattering is important \citep[cf.][]{2018ApJ...862..158E}. However, in order to highlight the effect of turbulence on the conductive term, here we do not consider its possible effect on the nonthermal electron transport, reserving such a study to a future work. \fp\ numerically solves the Fokker-Planck equation including Coulomb collisions and Ohmic losses associated with the return current electric field, on an energy grid with 300 logarithmically-spaced cells ranging from 0.1~keV to 10~MeV, using the method described in \citet{2020ApJ...902...16A}. As configured, \fp\ does not include the effects of runaway electrons, accelerated out of the ambient plasma by the return current electric field, on the transport of beam electrons. These can be important when the electron beam flux is high, but in this work the flux is sufficiently low that runaways likely only comprise a small fraction of the return current \citep{2021ApJ...917...74A}. Turbulence, which causes suppressed thermal conduction, also produces increased electrical resistivity, causing larger return current electric fields, but also a corresponding increase in the Dreicer field \citep{1959PhRv..115..238D}. Thus, we expect that the fraction of runways in the return current is not directly affected by the level of turbulence present.

\section{Modeling of the Atmospheric Response}\label{sec:discuss}

To study how suppressed heat flux affects the evolution of flaring loops, we have chosen six sets of the parameters $R$ and $\alpha$, and performed flare simulations for each. These parameter sets are listed in Table~\ref{tab:params}, where the first column indicates labels by which we refer to the models. For comparison, we also have performed a simulation of the unsuppressed case that implements thermal conduction using the standard Spitzer heat flux (Equation~(\ref{eqn:spitzer})) and is labeled ``Spitzer.'' To best compare with the Spitzer theory, our ``Spitzer'' model simply implements the heat flux using Equation~\ref{eqn:spitzer} and does not limit it by the electron free-streaming rate as was done in most previous \radyn\ simulations. The conductive flux suppression factor, $s$, is computed by integrating $Q \lambda_{ei}$ over $z/\lambda_{ei}$ (Figure~\ref{fig:kernel}), and is the ratio $F_c/F_S$ that would be obtained if $F_S$ does not appreciably vary over $z$, that is, when non-local effects are small. Of course, in these simulations non-locality can be important and the actual conduction suppression can deviate from $s$. 

The particular values of $R$ and $\alpha$ listed in Table~\ref{tab:params} were chosen to span a wide range of thermal conduction suppression, with $s$ ranging from $0.05$ to $1$. The most suppressed models (R1.0A0.9 and R1.0A1.7) have narrow kernels (see Figure~\ref{fig:kernel}) and demonstrate cases where turbulence dominates over non-locality. The opposite extreme, where non-locality dominates over turbulence, is demonstrated with the model R10.0A9.0, which has a relatively broad kernel. We have chosen two parameter sets that have $\alpha = 4.0$ (R1.0A4.0 and R2.0A4.0), since these have the interesting property of producing kappa-type velocity distribution functions (see Equation~(\ref{eqn:fw}) and the discussion in \S\ref{sec:profiles}). These have moderate suppression, with $s$ equal to 0.5 and 0.33, respectively. Finally, we have chosen the model, R13.5A6.0, which has a similar suppression factor, $s = 0.33$, as the R2.0A4.0 model, but since $\alpha = 6$ produces very broadened Maxwellian velocity distributions (again see Equation~(\ref{eqn:fw})).

\begin{deluxetable}{lcccccc}
\tabletypesize{\footnotesize}
\tablecaption{Label, $R$, $\alpha$, Suppression Factor $s$, Oscillation Period, Temperature and Density Relaxation Times\label{tab:params}}
\tablehead{\colhead{Label} & \colhead{R} & \colhead{$\alpha$} & \colhead{$s$} & \colhead{Period (s)} & \colhead{$\Delta t_T$ (s)} & \colhead{$\Delta t_n$ (s)}}
\startdata
  Spitzer &   N/A &   N/A &  1.00 &  55.8 &   402 &   688 \\
 R1.0A0.9 &   1.0 &   0.9 &  0.05 &  34.2 &  1556 &  2117 \\
 R1.0A1.7 &   1.0 &   1.7 &  0.10 &  39.0 &  1143 &  1625 \\
 R1.0A4.0 &   1.0 &   4.0 &  0.50 &  49.3 &   552 &   884 \\
 R2.0A4.0 &   2.0 &   4.0 &  0.33 &  46.4 &   661 &  1027 \\
R10.0A9.0 &  10.0 &   9.0 &  0.90 &  51.3 &   422 &   722 \\
R13.5A6.0 &  13.5 &   6.0 &  0.33 &  46.3 &   662 &  1030 \\
\enddata
\end{deluxetable}

\subsection{Thermal evolution}\label{sec:tempevol}
\begin{figure*}
\plotone{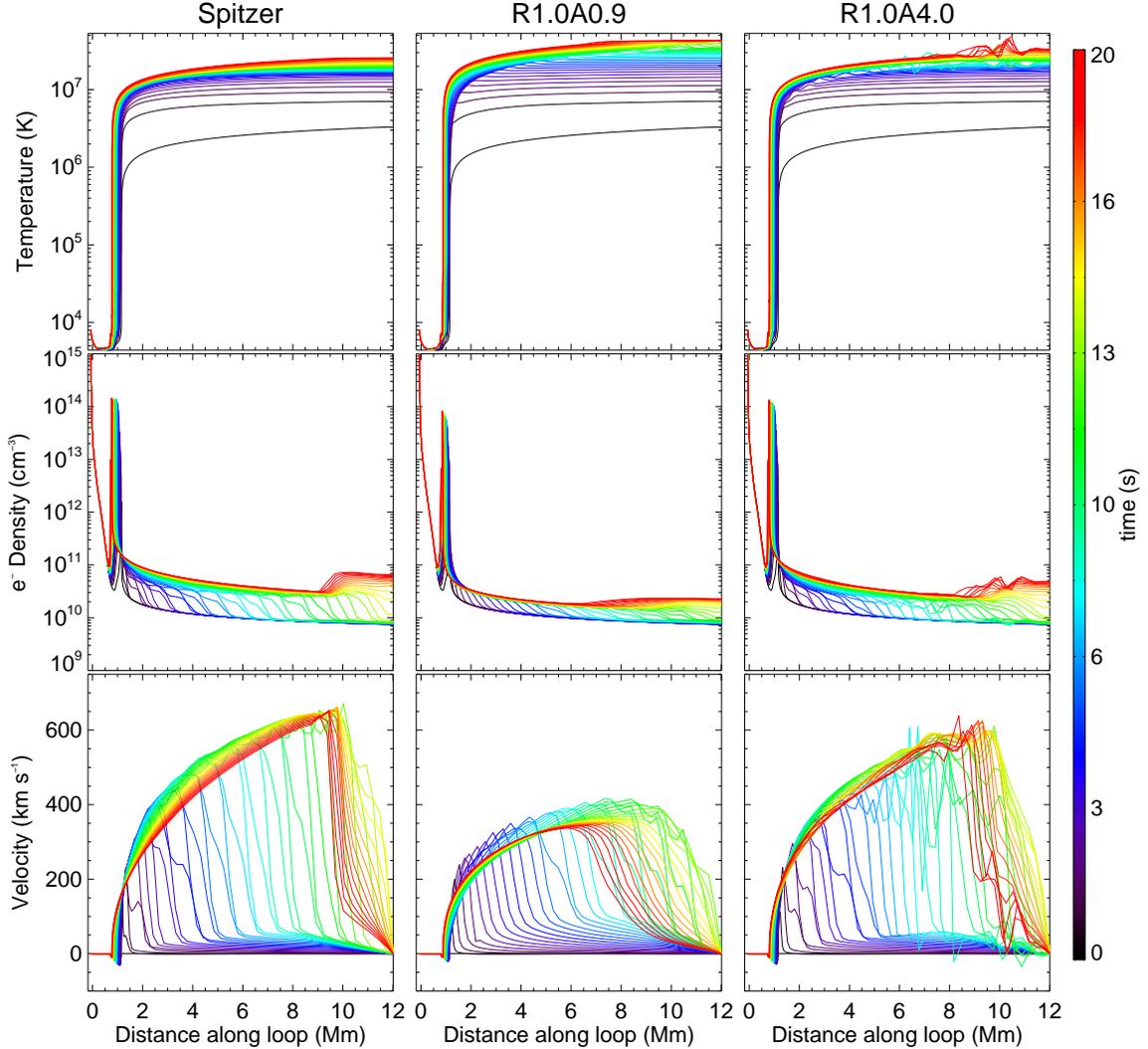}
\caption{The temporal evolution of the temperature (top row), electron density (middle row), and gas velocity (bottom row) for the Spitzer (left column), R1.0A0.9 (middle column), and R1.0A4.0 (right column) simulations. The line color represents the time from initial beam injection as indicated from the color bar.\label{fig:loopevol}}
\end{figure*}

Figure~\ref{fig:loopevol} shows the evolution of the temperature, electron density, and gas velocity for the first 20~s of the R1.0A0.9, R1.0A4.0, and Spitzer simulations. We chose to show these three models because they represent the most, middle and least suppressed cases. The R1.0A0.9 and R1.0A4.0 simulations reach hotter temperatures but have lower coronal densities than the Spitzer simulation. This is because the reduced heat flux limits how quickly heat can be moved from the corona into the chromosphere causing it to ``bottle-up'' in the corona and limiting the responding rate of chromospheric evaporation. The transition region forms at a height where the density is such that the radiative losses balance the heat flux divergence. A reduced heat flux, therefore, causes the transition region to form higher in the loop. The flaring transition region (at $t = 20$~s) is at 0.77~Mm, 0.79~Mm, and 0.87~Mm, for the Spitzer,  R1.0A4.0, and R1.0A0.9 simulations, respectively. 

\begin{figure}
\plotone{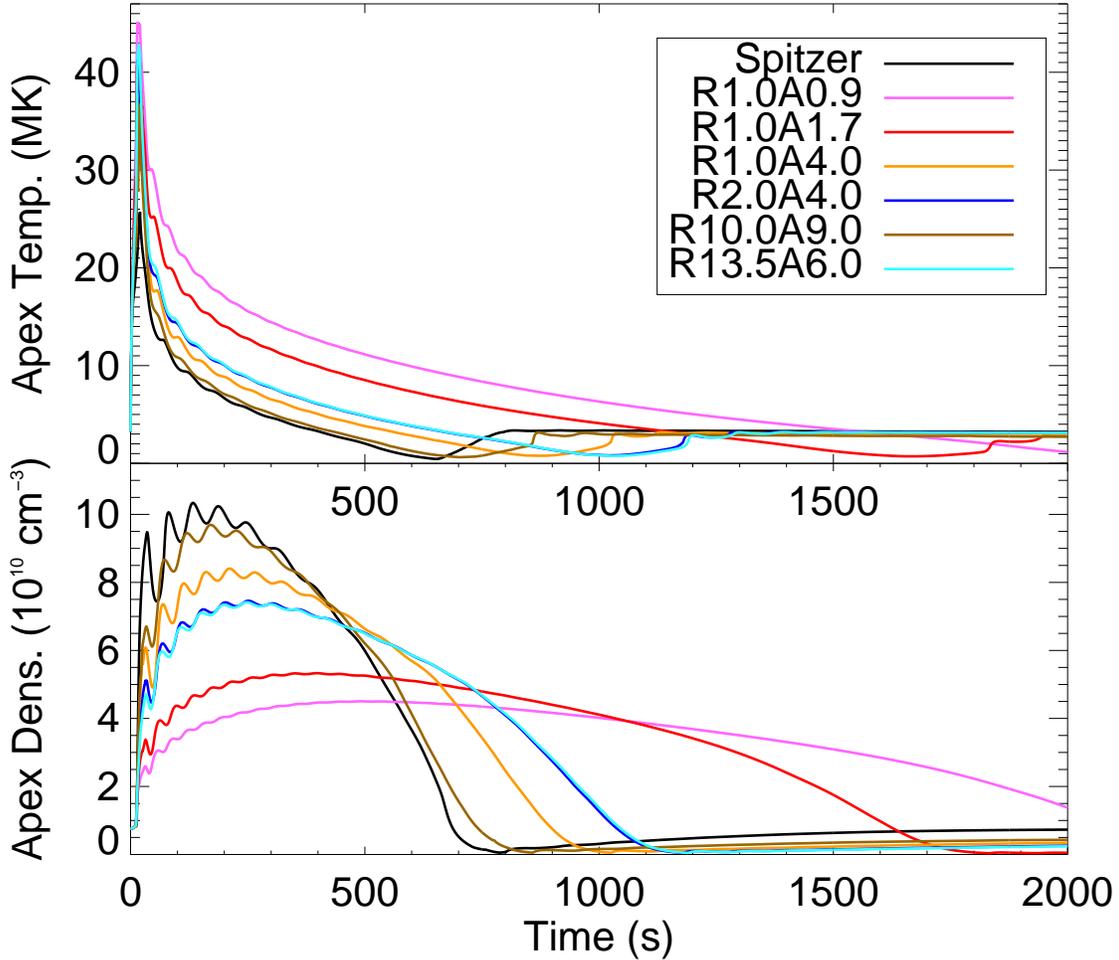}
\caption{The apex temperature (top) and density (bottom) as a function of time for the Spitzer (black), R1.0A0.9 (magenta), R1.0A1.7 (red), R1.0A4.0 (orange), R2.0A4.0 (dark blue), R10.0A9.0 (brown),  and R13.5A6.0 (cyan) simulations.\label{fig:apex}}
\end{figure}

In Figure~\ref{fig:apex}, we plot the temporal evolution of the apex temperature and density for each of the flare simulations. Oscillations are apparent in each case, and their periods are written in the fifth column of Table~\ref{tab:params}. Models with more reduced heat flux have shorter oscillation periods. Since their coronal temperatures are hotter, they have faster sound speeds, so pulses travel through the loop in shorter times. 

\begin{figure}
\plotone{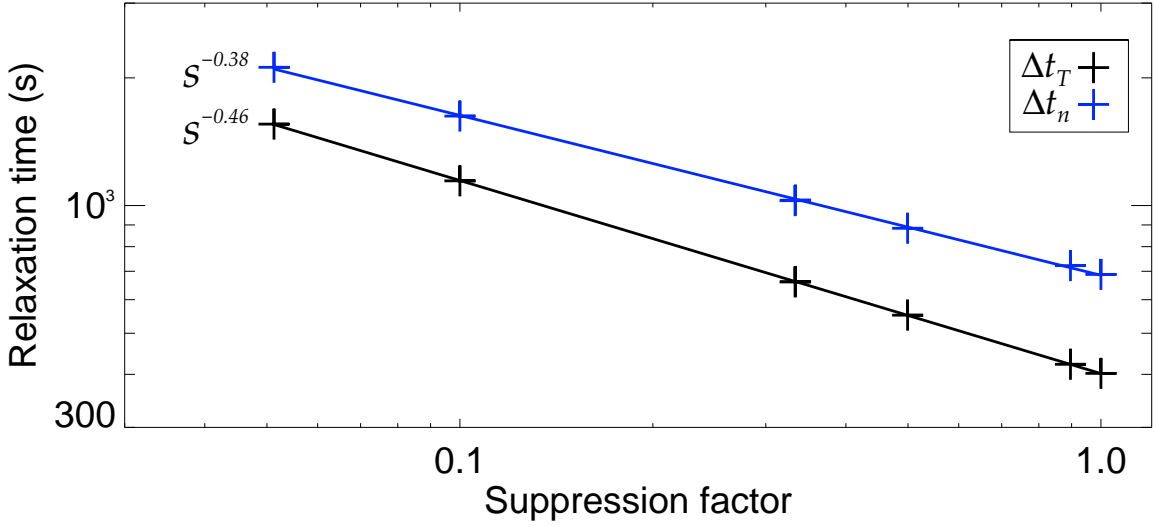}
\caption{The temperature (black plus signs) and density (blue plus signs) relaxation times as a function of suppression factor, $s$, for the seven models in Table~\ref{tab:params}. The solid lines indicate best fit power laws with slopes of $-0.46$ and $-0.38$ for the temperature and density relaxation times, respectively.\label{fig:relaxtime}}
\end{figure}

We define the temperature ($\Delta t_T$) and density ($\Delta t_n$) relaxation times as the time intervals required for the model looptops to cool and drain to their preflare values. These are listed in the sixth and seventh columns of Table~\ref{tab:params}. More suppressed models have longer relaxation times. As discussed in EB18, turbulent suppression of conduction may aid in explaining longer than predicted cooling times observed for numerous flares \citep{2013ApJ...778...68R}. The temperature and density relaxation times as a function of $s$ are well-fitted by power-laws of the suppression factor $s$: $\Delta t_{T,n} \propto s^{-\beta_{T,n}}$, with indices $\beta_T = 0.46$ and $\beta_n = 0.38$, respectively, as shown in Figure~\ref{fig:relaxtime}. 

\subsection{Spectral line profiles}\label{sec:profiles}
Using \radyn's predicted temporal evolution of the temperatures and electron and hydrogen densities combined with atomic line contribution functions obtained from the CHIANTI package \citep[v.$10.0.1$;][]{1997A&AS..125..149D,2021ApJ...909...38D}, we have constructed line intensities along our model loop for 14 optically-thin transition region and coronal atomic lines presented in MD09 (listed in their Table~1), with the exception of the \ionwl{He}{2}{256} line, which is likely optically-thick and not well-suited to CHIANTI's coronal approximation. We use CHIANTI's ionization equilibrium table and the abundance table from \citet{2012ApJ...755...33S}, but we note that since our primary purpose in this section is to compare profile shapes and Doppler velocities, differences in abundance tables are not critical to our study. 

In a thermal non-turbulent plasma, the velocity distribution function, $f(v)$, is a Maxwellian with width given by the thermal velocity. With this assumption, line profiles will be Gaussians. However, the presence of turbulence causes the velocity distribution function to deviate from a Maxwellian resulting in non-Gaussian line profiles, with widths that are typically wider than than the thermal width. \citet{2014ApJ...796..142B} solved for the velocity distribution function in a turbulent environment using a steady-state approximation to the Fokker-Planck equation. They found a distribution function of the form (their Equations (7)-(10)),

\begin{linenomath*} \begin{equation}
f(v) = A \, e^{-U(v)} \,\,\, ,
\label{eqn:distf}
\end{equation} \end{linenomath*}
with $U(v)$ the solution of

\begin{linenomath*} \begin{equation}
\frac{d U}{d v} = \left( \frac{v_{te}^2}{2 v} + \frac{v^2 D_{turb}}{\Gamma} \right)^{-1} \,\,\, ,
\label{eqn:uprime}
\end{equation} \end{linenomath*}
where $\Gamma = v_{te}^4/\lambda_{ei}$. The turbulent diffusion coefficient, $D_{turb}$, is given by

\begin{linenomath*} \begin{equation}
D_{turb} = \frac{v^3}{\lambda_T(v)} = \frac{v^{3-\alpha} v_{te}^\alpha}{\lambda_{T0}} \,\,\, ,
\end{equation} \end{linenomath*}
where we have used the expression for the turbulent mean free path in Equation~(\ref{eqn:vturb}). Substituting this into Equation~(\ref{eqn:uprime}) gives

\begin{linenomath*} \begin{equation}
\frac{d U}{d v} = \left( \frac{v_{te}^2}{2 v} + R \, \frac{v^{5-\alpha}}{v_{te}^{4-\alpha}} \right)^{-1} = \frac{2 \, v \, v_{te}^{4-\alpha}}{v_{te}^{6-\alpha} + 2 R v^{6-\alpha}} \,\,\, .
\end{equation} \end{linenomath*}
In terms of the dimensionless speed $w = v/v_{te}$,

\begin{linenomath*} \begin{equation}
\frac{d U}{d w} = \frac{2 w}{1 + 2 R w^{6 - \alpha}} \,\,\, .
\end{equation} \end{linenomath*}
Finally, we obtain the distribution function by integrating this and substituting into Equation~(\ref{eqn:distf})

\begin{linenomath*} \begin{equation}
f(w) = A_0 \exp \left[-\int_0^w \frac{2 x dx}{1 + 2 R x^{6-\alpha}} \right] \,\,\, ,
\label{eqn:fw}
\end{equation} \end{linenomath*}
where the coefficient $A_0$ is chosen to normalize $f(w)$ to the local density. From this form of $f(w)$, we note:

\begin{enumerate}
\item In the no-turbulence limit, i.e., when $R=0$, $f(w)$ regains its thermal Maxwellian form, as expected;
\item For the case with $\alpha = 6$, $f(w)$ is also a Maxwellian, but is now broadened by a factor of $\sqrt{1+2 R}$. This result follows from the fact that the collisional and turbulent terms appearing in the Fokker-Planck equation have the same velocity dependence (i.e., $v^{-3}$). Thus, as previously noted by \cite{2014ApJ...796..142B}, the presence of a Maxwellian distribution function is {\it not} exclusive to the collisions-only situation. Caution must therefore be observed in inferring the local temperature from the width of an observed Gaussian line profile;
\item For the case with $\alpha = 4$, $D_{turb}$ has a velocity dependence of $v^{-1}$. In this case, $f(w)$ becomes a kappa distribution of the first kind, as discussed in Section~3 of \citet{2014ApJ...796..142B}, with $\kappa = 1/2 R$;
\item For $\alpha < 4$, the dependence of $D_{turb}$ on $v$ is such that, as $w \to \infty$, $f(w)$ behaves as an exponential of a negative power of $w$. For example, with $\alpha = 2$, $f(w) \to A e^{1/(2 R w^2)}$ , while for $\alpha = 0$, $f(w) \to A e^{1/(4 R w^4)}$. For large $w$, these approach an asymptotic value, $A$, but this steady-state is only reached as $t \to \infty$ \citep{PhysRevE.72.061106}. So, to normalize $f(w)$ for $\alpha < 4$, we truncate $f(w)$ at 10 Doppler widths and then subtract off the asymptotic value to ensure that the line profiles go to zero for large $w$.
\end{enumerate}

For a given set of $R$ and $\alpha$, line profiles are obtained by numerically calculating the integral in Equation~(\ref{eqn:fw}) and are normalized so that the integral over wavelength is the total line intensity. The line profiles are then Doppler shifted by the bulk velocity along the loop axis projected along an observer's line-of-sight. MD09 measured line-of-sight Doppler velocities for several lines in EIS pixels that corresponded to the location of the hard X-ray footpoint as observed by RHESSI (see their Figures~3--5). To better enable comparisons with this flare observation, we orient our model loop footpoint at the same location on the solar disk, which is $(X, Y) = (550", -94")$. At the footpoint, we assume that the loop is oriented along the vertical axis. Then using the WCS conversion routine, \texttt{WCS\_CONV\_HPC\_HCC}, the projection between an observer at earth and the vertical at the footpoint is readily determined to be $\cos(35.6^{\circ}) = 0.81$. We use this factor as the line-of-sight projection of bulk flows along our model loop. The EIS observation of this flare used a slit with angular width of 2", spectral instrumental full width at half maximum ($W_{inst}$) of approximately 0.056~\AA\  \citep{2009ApJ...691L..99H}, and a temporal integration time of 10~s, so to enable comparison to the EIS observation, we bin the predicted emissions from our model loops into 2" bins and integrate them over 10~s intervals. The modeled line profiles are convolved with a Gaussian with width given by $W_{inst}$ to simulate instrumental broadening. 

\begin{figure*}
\plotone{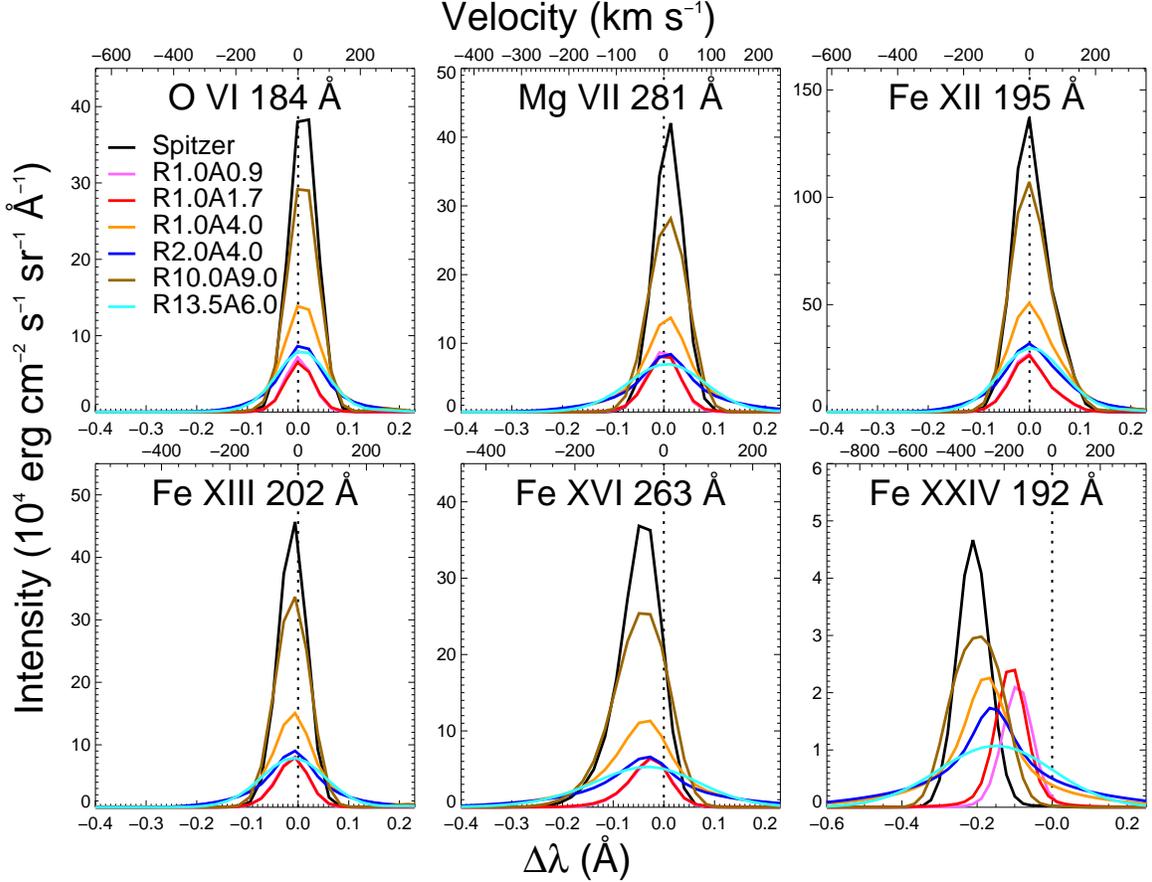}
\caption{Line profiles averaged over the first 10~s of the Spitzer (black), R1.0A0.9 (magenta), R1.0A1.7 (red), R1.0A4.0 (orange), R2.0A4.0 (dark blue), R10.0A9.0 (brown),  and R13.5A6.0 (cyan) simulations for the \ionwl{0}{6}{184} (top left), \ionwl{Mg}{7}{281} (top middle), \ionwl{Fe}{12}{195} (top right), \ionwl{Fe}{13}{202} (bottom left), \ionwl{Fe}{16}{263} (bottom middle), and \ionwl{Fe}{24}{192} (bottom right) atomic lines. In each panel, the bottom and top horizontal axes are the wavelength shifts from rest (\AA) and corresponding Doppler velocity (km s$^{-1}$), respectively. Negative velocity is upward. The vertical black dotted lines indicate zero shift.\label{fig:lp}}
\end{figure*}

We have produced line profiles for the 14 lines listed in Table~\ref{tab:lines}. As examples, we show line profiles for the first 10~s interval of each of our simulations for the \ionwl{0}{6}{184}, \ionwl{Fe}{12}{195}, \ionwl{Fe}{16}{263}, and \ionwl{Fe}{24}{192} lines in Figure~\ref{fig:lp}. Since a reduced heat flux drives slower chromospheric evaporation with a less dense corona, for each of these lines, models with more suppressed heat flux are less intense and exhibit weaker Doppler shifts. For example, comparing our most suppressed model (R1.0A0.9) with the Spitzer model, the \ionwl{Fe}{24}{192} line has integrated intensities of $\sn{2.2}{3}$ and $\sn{5.0}{3}~$erg cm$^{-2}$ s$^{-1}$ sr$^{-1}$ and Doppler shifts of $-144$ and $-328$~km s$^{-1}$, respectively. 

Characteristics of the line profiles are summarized in Table~\ref{tab:lines}. The third column is the ion formation temperature from CHIANTI~v.$10.0.1$. These differ slightly for a few lines compared to M11, which used CHIANTI~v.$6.0.1$. The fourth and fifth columns are the Doppler speed (with uncertainties) and nonthermal velocities determined in MD09/M11 (see Table~1 of M11). The left and right columns under each model heading are the Doppler and nonthermal velocities, respectively. The Doppler shifts are calculated from the line profile centroid. The nonthermal line velocities are found by measuring the full width at half maximum, $W$, and subtracting the contributions from thermal and instrumental broadening. Specifically, we measure the nonthermal velocity, $v_{NT}$, using the expression

\begin{linenomath*}\begin{equation}
v_{NT} = \left[\frac{c^2}{4 \, \lambda_c^2 \, \ln 2} \left(W^2 - W_{inst}^2 \right) - \frac{2 k T_{max}}{m_i} \right]^\frac{1}{2},
\end{equation}\end{linenomath*}
where $c$ is the speed of light, $\lambda_c$ is the line center wavelength, $T_{max}$ is the temperature at the peak of the contribution function (from CHIANTI), and $m_i$ is the ion mass.

\begin{deluxetable*}{lcc|cc|cc|cc|cc|cc|cc|cc|cc}
\tabletypesize{\scriptsize}
\tablecaption{Doppler ($v$) and nonthermal ($v_{NT}$) velocities measured in several lines from MD09/M11 and from our models\label{tab:lines}}
\tablehead{
\colhead{Ion} & \colhead{$\lambda_0$} & \colhead{$T_{max}$} & \multicolumn{2}{c}{MD09/M11} & \multicolumn{14}{c}{Model} \\
\cline{6-19}
\colhead{}  & \colhead{} & \colhead{}  &  \colhead{} & \colhead{} & \multicolumn{2}{c}{Spitzer} & \multicolumn{2}{c}{R1.0A0.9} 
  & \multicolumn{2}{c}{R1.0A1.7} & \multicolumn{2}{c}{R1.0A4.0} & \multicolumn{2}{c}{R2.0A4.0} & \multicolumn{2}{c}{R10.0A9.0} & \multicolumn{2}{c}{R13.5A6.0} \\[-8pt]
\colhead{} & \colhead{} & \colhead{} & \colhead{$v$} & \colhead{$v_{NT}$} & \colhead{$v$} & \colhead{$v_{NT}$}& \colhead{$v$} & \colhead{$v_{NT}$}  & \colhead{$v$} & \colhead{$v_{NT}$}  
& \colhead{$v$} & \colhead{$v_{NT}$} & \colhead{$v$} & \colhead{$v_{NT}$} & \colhead{$v$} & \colhead{$v_{NT}$} & \colhead{$v$} & \colhead{$v_{NT}$} \\[-8pt]
\colhead{} & \colhead{(\AA)} & \colhead{(MK)\tablenotemark{\scriptsize{a}}} & \multicolumn{16}{c}{(km s$^{-1}$)} }
\startdata
  \ion{O}{6} & 184.12  & 0.3   & $60   \pm 14$ & 68  & $16$ &  $33$ &  $2$ &   $26$ &  $4$ &   $32$ &  $14$ &  $65$ &  $12$ &  $78$ &  $15$ &  $46$ &  $12$ &  $103$ \\
 \ion{Mg}{6} & 268.99  & 0.4   & $51   \pm 15$ & 71  & $15$ &  $15$ &  $2$ &   $25$ &  $4$ &   $30$ &  $12$ &  $46$ &  $10$ &  $59$ &  $14$ &  $35$ &  $10$ &  $90$ \\ 
 \ion{Mg}{7} & 280.74  & 0.6   & $53   \pm 13$ & 64  & $11$ &  $16$ &  $-1$ &  $28$ &  $0$ &   $32$ &  $7$ &   $49$ &  $6$ &   $64$ &  $9$ &   $42$ &  $6$ &   $107$ \\
 \ion{Fe}{8} & 185.21  & 0.6   & $33   \pm 17$ & 74  & $16$ &  $28$ &  $-1$ &  $33$ &  $1$ &   $37$ &  $11$ &  $56$ &  $9$ &   $68$ &  $12$ &  $34$ &  $9$ &   $72$ \\ 
\ion{Fe}{10} & 184.54  & 1.1   & $35   \pm 16$ & 97  & $-2$ &  $22$ &  $-9$ &  $33$ &  $-9$ &  $35$ &  $-2$ &  $56$ &  $-3$ &  $72$ &  $-2$ &  $40$ &  $-3$ &  $95$ \\ 
\ion{Fe}{11} & 188.22  & 1.3   & $43   \pm 15$ & 60  & $-5$ &  $33$ &  $-6$ &  $49$ &  $-5$ &  $48$ &  $4$ &   $71$ &  $10$ &  $96$ &  $-2$ &  $54$ &  $29$ &  $153$ \\
\ion{Fe}{12} & 195.12  & 1.6   & $28   \pm 17$ & 81  & $-1$ &  $43$ &  $-5$ &  $52$ &  $-3$ &  $51$ &  $5$ &   $83$ &  $7$ &   $101$ & $1$ &   $62$ &  $7$ &   $125$ \\
\ion{Fe}{13} & 202.04  & 1.8   & $-18  \pm 14$ & 54  & $-15$ & $24$ &  $-16$ & $38$ &  $-15$ & $36$ &  $-15$ & $61$ &  $-14$ & $79$ &  $-16$ & $48$ &  $-14$ & $121$ \\
\ion{Fe}{14} & 274.20  & 2.0   & $-22  \pm 12$ & 58  & $-22$ & $24$ &  $-17$ & $35$ &  $-16$ & $33$ &  $-20$ & $57$ &  $-19$ & $74$ &  $-22$ & $53$ &  $-19$ & $130$ \\
\ion{Fe}{15} & 284.16  & 2.2   & $-32  \pm 8 $ & 73  & $-31$ & $28$ &  $-21$ & $35$ &  $-21$ & $32$ &  $-28$ & $61$ &  $-27$ & $76$ &  $-30$ & $60$ &  $-26$ & $141$ \\
\ion{Fe}{16} & 262.98  & 2.8   & $-39  \pm 20$ & 48  & $-48$ & $43$ &  $-29$ & $39$ &  $-31$ & $37$ &  $-46$ & $77$ &  $-43$ & $92$ &  $-48$ & $73$ &  $-40$ & $163$ \\
\ion{Fe}{17} & 269.42  & 5.6   & $-69  \pm 18$ & 78  & $-99$ & $70$ &  $-50$ & $44$ &  $-59$ & $49$ &  $-90$ & $102$ & $-84$ & $121$ & $-94$ & $97$ &  $-79$ & $209$ \\
\ion{Fe}{23} & 263.77  & 14.1  & $< -230 \pm 32$ & 122 & $-292$ &$55$ &  $-132$ &$31$ &  $-158$ &$28$ &  $-242$ &$108$ & $-220$ &$133$ & $-275$ &$134$ & $-217$ &$313$ \\
\ion{Fe}{24} & 192.03  & 17.8  & $< -257 \pm 28$ & 105 & $-328$ &$31$ &  $-144$ &$10$ &  $-172$ &$20$ &  $-273$ &$112$ & $-247$ &$143$ & $-311$ &$136$ & $-235$ &$324$ \\
\enddata
\tablenotetext{a}{Formation temperatures are from CHIANTI v.$10.0.1$~\citep{2021ApJ...909...38D} and differ slightly from those listed in M11, which used CHIANTI v.$6.0.1$.}
\end{deluxetable*}

\subsection{Doppler Velocities}
\begin{figure}
\plottwo{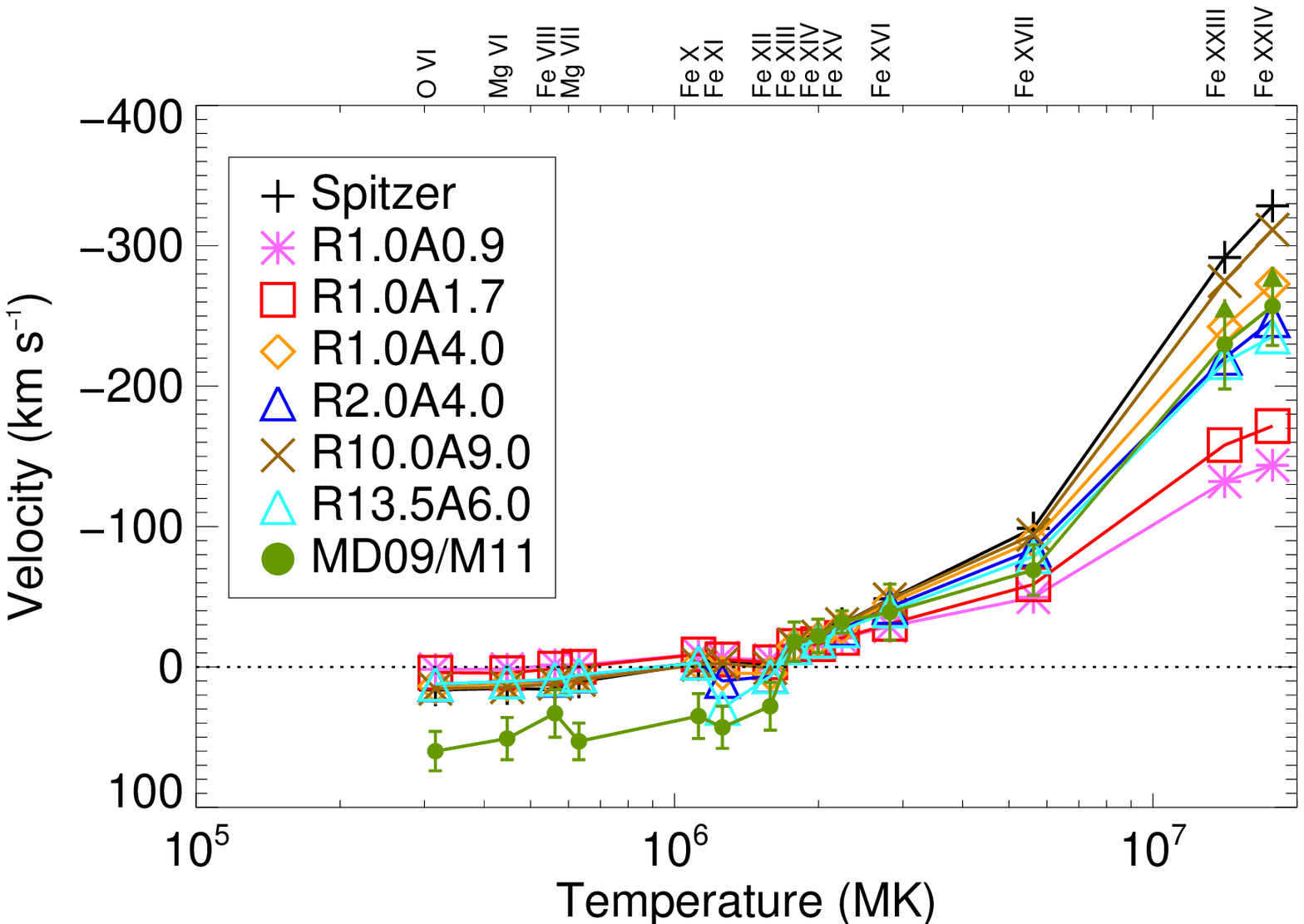}{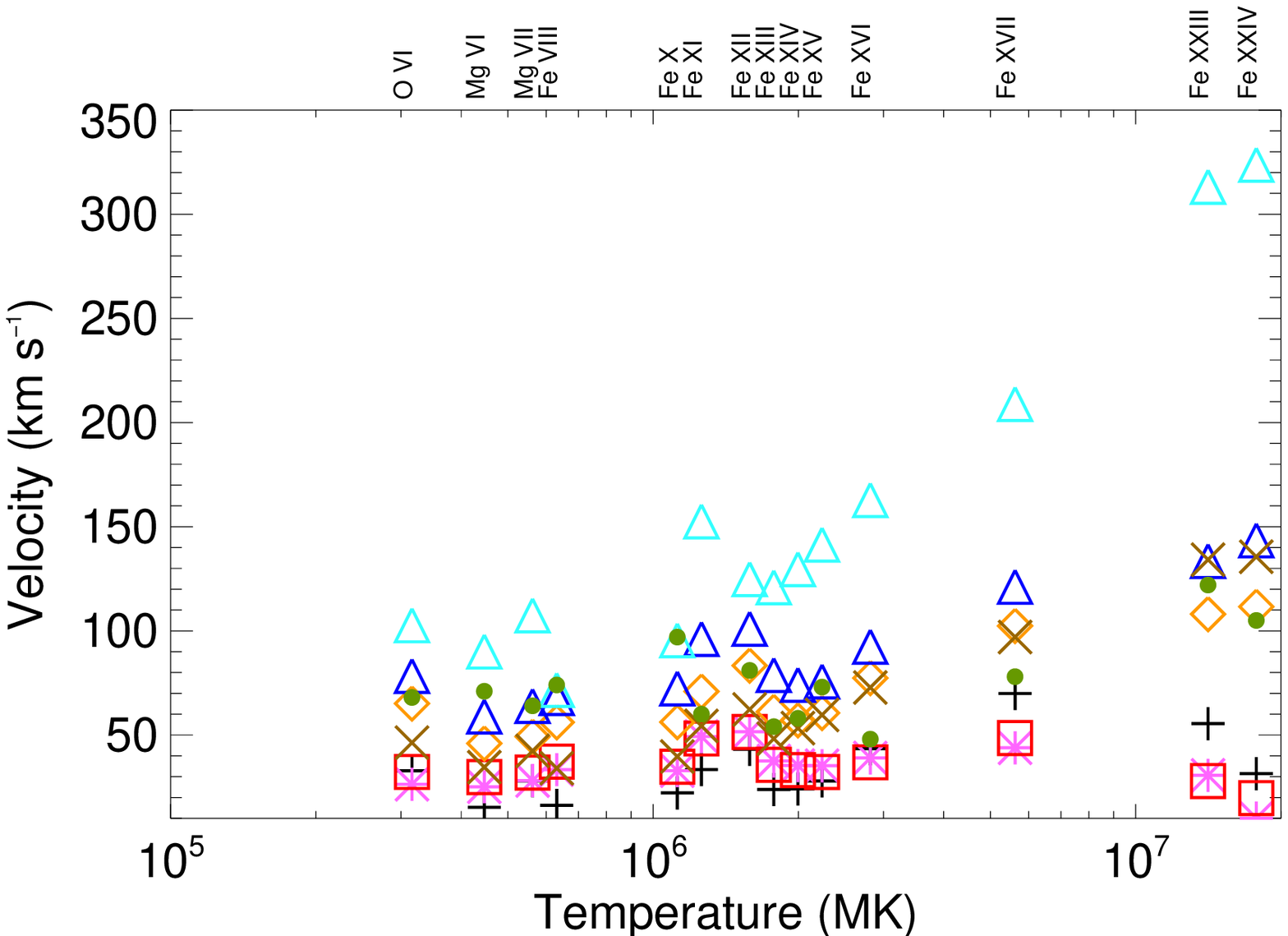}
\caption{Doppler (left panel) and nonthermal velocities (right panel) measured from EIS observations of several atomic lines by MD09/M11(green circles) compared with predicted  velocities from the Spitzer (black plus signs), R1.0A0.9 (magenta asterisks), R1.0A1.7 (red squares), R1.0A4.0 (orange diamonds), R2.0A4.0 (dark blue triangles), R10.0A9.0 (brown $\times$'s),  and R13.5A6.0 (cyan triangles) simulations. The lines are listed along the top of the plots and the horizontal axis is the expected temperature of formation for the line. In the left plot, negative velocity is upward and the horizontal black dotted line indicates zero velocity.}\label{fig:dopspeed}
\end{figure}

To better visualize how the Doppler velocities compare to the MD09/M11 observations, in the left panel of Figure~\ref{fig:dopspeed} we show the Doppler velocities predicted from each of our simulations as a function of line formation temperature (see also Figure~5 of MD09). All of our simulations produce downflows in cooler lines (chromospheric condensations) and upflows in hotter lines (chromospheric evaporations). Similar to the to MD09/M11 observations, the R1.0A4.0, R2.0A4.0, and R13.5A6 simulations predict the transition between upflows and downflows occurs between the \ion{Fe}{12} ($T_{max} = 1.6$~MK) and \ion{Fe}{13} ($T_{max} = 1.8$~MK) lines. However, the simulations predict downflows that are significantly slower than the observations. For example, we find flows less than $16$ km s$^{-1}$ in the \ion{O}{6} line compared to the observed $(60 \pm 14)$~km~s$^{-1}$. The only exception is the \ionwl{Fe}{11}{188} line in the R13.5A6.0 model, which appears to have a faster downflow (29~km~s$^{-1}$) compared to the other models. However, that is a result of its very broad line profile merging with a neighboring (redward) \ion{Fe}{11} line and not a real redshift. The too slow downflows could be an indication that the preflare model chromosphere is relatively overdense compared to the real Sun, or perhaps the energy flux density in the electron beam was underestimated. This is possible if the footpoint area is not fully resolved. 

The speed of the chromospheric evaporation is more directly related to the heat flux. The models with low suppression, Spitzer and R10.0A9.0, predict upflows that are faster than observed in the \ion{Fe}{16} and hotter lines. For the strongly suppressed models, the opposite is true. The R1.0A0.9 and R1.0A1.7 models predict $-144$ and $-172$ compared to the observed $<-257$~km~s$^{-1}$ for the \ionwl{Fe}{24}{192}. Interestingly, the moderately suppressed models, R1.0A4.0, R2.0A4.0, and R13.5R6.0 predict upflow velocities that are within the MD09/M11 uncertainties. Therefore, these have plausible levels of thermal conduction suppression. 

\subsection{Nonthermal Velocities}
The nonthermal velocities listed in Table~\ref{tab:lines} are plotted in the right panel of Figure~\ref{fig:dopspeed}. Even though the Spitzer simulation has no turbulence, it shows a small level of excess broadening (seventh column of Table~\ref{tab:lines}), resulting from the superposition of a distribution of Doppler shifts occurring within the temporal and spatial bins, as speculated by M11 and discussed in \citet{2019ApJ...879L..17P} and \citet{2020ApJ...900...18K} in the context of \ion{Fe}{21}. But this small amount of broadening is only comparable to the M11 observations for the \ionwl{Fe}{16}{263} and \ionwl{Fe}{17}{269} lines. For other lines, especially the very hot \ion{Fe}{23} and \ion{Fe}{24} lines, the Spitzer simulation is unable to produce sufficiently broad line profiles. For example, M11 observed $v_{NT}$ = 122~km~s$^{-1}$ in \ionwl{Fe}{23}{264} line compared to $55$ km s$^{-1}$ predicted by the Spitzer simulation. The highly suppressed models, R1.0A0.9 and R1.0A1.7, also predict line profiles that are much narrower than observed. The R13.5A6 model has a broadening factor of $\sqrt{1 + 2(13.5)} = 5.3$ and produces line profiles that are much broader than in M11, e.g., 313~km~s$^{-1}$ in the \ion{Fe}{23} line. M11 did not report the uncertainties in the nonthermal line widths, but assuming that they are comparable to the reported Doppler velocity uncertainties, we find that R1.0A4.0, R2.0A4.0, and R10.0A9.0 produce nonthermal broadenings that are similar to the observations. Of these, R1.0A4.0 has the smallest squared residuals. As described in the previous section, this model also produces Doppler velocities that match well with the observations. It is, therefore, plausible that the nonthermal line widths and Doppler velocities observed by MD09/M11 indicate a moderate level of turbulence produced during that flare. Interestingly, in our best fit model, since $\alpha = 4$, the velocity distribution function is characterized by a kappa distribution, with a rather low value of $\kappa = 1/2R = 1/2$. In the analysis of  other flares, \citet{2018ApJ...864...63P} and \citet{2016A&A...590A..99J} also found kappa distributions (with significantly higher values of $\kappa \simeq (2 - 3)$) best fit line profiles, and \citet{2018ApJ...864...63P} concluded that turbulence is a likely cause.

\section{Summary and Conclusions}\label{sec:conc}
We have used the \radyn\ flare modeling code to study how the effects of turbulence and non-locality on conductive heat transport affect flare radiation hydrodynamics. Specifically, we have implemented the EB18 thermal conduction model into \radyn\ and performed a parameter study, varying $R$ and $\alpha$, to understand how the suppressed heat flux affects the speed of chromospheric evaporation and the time scales of flare cooling and draining. We have studied seven sets of parameters with suppression factor ranging from $0.05$ to $1.0$ (i.e., unsuppressed). We used \radyn\ to simulate an injection of a beam of nonthermal electrons and the subsequent heating. The same beam distribution was used for each simulation, and was chosen to match the power-law inferred from a RHESSI observation of a C-class flare by MD09. Our key findings are summarized below:

\begin{enumerate}
\item The transition region forms at a lower density in more highly suppressed models, resulting in lower intensities in transition region and coronal spectral lines;
\item Models with more highly suppressed thermal conduction result in longer cooling ($\Delta t_T$) and draining ($\Delta t_n$) times. These are well-fitted by power-laws of the suppression factor $s$: $\Delta t_{T,n} \propto s^{-\beta_{T,n}}$ with indices $\beta_{T,n}$ equal to $0.46$ and $0.38$, respectively; 
\item The flaring chromospheric evaporation front is slower for models with more highly suppressed thermal conduction. The models R1.0A4.0, R2.0A4.0, and R13.5R6.0, which have $s = 0.5$, $0.33$, and $0.33$, respectively, produce upflow Doppler shifts in several lines that are comparable to observations of a C-class flare presented in MD09/M11;
\item Predicted nonthermal line widths, primarily due to turbulent broadening, from the R1.0A4.0, R2.0A4.0, and R10.0A9.0 models are comparable to the M11 observations;
\item Overall, the R1.0A4.0 model produces Doppler velocities and nonthermal line widths that best match the observations. This model has a moderate level of turbulent suppression with $s = 0.5$. Since $R=1.0$, the turbulent and collisional mean free paths are similar. In this model $\alpha = 4.0$, so the velocity distribution function is a kappa distribution of the first kind with a rather low value of $\kappa = 1/2R = 1/2$.
\end{enumerate}

This work represents a first step in understanding how turbulence affects the evolution of flare loops and the resulting emission. In future work, we will further improve upon our models by including the effects of turbulent diffusion on the transport of the nonthermal electrons and the effects of runaway electrons accelerated out of the ambient thermal distribution.

\acknowledgments
J.C.A. and G.S.K. acknowledge funding from NASA's Heliophysics Innovation Fund and Heliophysics Supporting Research program. Additionally, G.S.K. acknowledges funding via a NASA ROSES Early Career Investigator Award (Grant\# 80NSSC21K0460). A.G.E acknowledges funding from NASA Kentucky under NASA award number 80NSSC21M0362. This manuscript benefited from discussions held at a meeting of International Space Science Institute team: ``Interrogating Field-Aligned Solar Flare Models: Comparing, Contrasting and Improving,'' led by Dr.~G.~S.~Kerr and Dr.~V.~Polito.

\bibliography{hfpaper}

\end{document}